\documentclass[twocolumn]{aastex63}
\usepackage{graphicx}	
\usepackage{amsmath}	
\usepackage{amssymb}	
\usepackage{epstopdf}
\usepackage{multirow}
\accepted{April 19, 2022}
\submitjournal{ApJ}

\shorttitle{Extended X-ray Emission in Mrk 78}
\shortauthors{Fornasini et al.}
\graphicspath{{./}{figures/}}

\defcitealias{whittle04}{WW04}
\defcitealias{whittle05}{W05}
\defcitealias{fischer13}{F13}
\defcitealias{fischer11}{F11}
\defcitealias{revalski21}{R21}

\begin{document}

\title{Termination Shocks and the Extended X-ray Emission in MRK 78}

\newcommand\T{\rule{0pt}{2.6ex}}       
\newcommand\B{\rule[-1.2ex]{0pt}{0pt}} 

\correspondingauthor{F. M. Fornasini}
\email{ffornasini@stonehill.edu}

\author{Francesca M. Fornasini}
\affiliation{Center for Astrophysics $\vert$ Harvard \& Smithsonian, 60 Garden Street, Cambridge, MA 02138, USA}
\affiliation{Stonehill College, 320 Washington Street, Easton, MA 02357, USA}
\author{Martin Elvis}
\affiliation{Center for Astrophysics $\vert$ Harvard \& Smithsonian, 60 Garden Street, Cambridge, MA 02138, USA}
\author{W. Peter Maksym}
\affiliation{Center for Astrophysics $\vert$ Harvard \& Smithsonian, 60 Garden Street, Cambridge, MA 02138, USA}
\author{Giuseppina Fabbiano}
\affiliation{Center for Astrophysics $\vert$ Harvard \& Smithsonian, 60 Garden Street, Cambridge, MA 02138, USA}
\author{Thaisa Storchi Bergmann}
\affiliation{Instituto de Fisica - UFRGS, Campus do Vale, CP 15051 91501-970 Porto Alegre - RS - Brasil}
\author{Poshak Gandhi}
\affiliation{University of Southampton, University Road, 
Southampton SO17 1BJ, United Kingdom}
\author{Mark Whittle}
\affiliation{Department of Astronomy, University of Virginia, Charlottesville, VA 22903}




\begin{abstract}
Sub-arcsecond imaging of the X-ray emission in the type 2 AGN Mrk 78 with {\em Chandra} shows complex structure with spectral variations on scales from $\sim$200 pc to $\sim$ 2 kpc. Overall the X-ray emission is aligned E-W with the radio (3.6 cm) and narrow emission line region as mapped in [OIII], with a marked E-W asymmetry. The Eastern X-ray emission is mostly in a compact knot coincident with the location where the radio source is deflected, while the Western X-ray emission forms a loop or shell $\sim$2 kpc from the nucleus with radius $\sim$0.7 kpc. There is suggestive evidence of shocks in both the Eastern knot and the Western arc.  Both these positions coincide with large changes in the velocities of the [OIII] outflow.  We discuss possible reasons why the X-ray shocks on the Western side occur $\sim1$ kpc farther out than on the Eastern side.  We estimate that the thermal energy injected by the shocks into the interstellar medium corresponds to $0.05-0.6$\% of the AGN bolometric luminosity.

\end{abstract}

\keywords{}



\section{Introduction} 
\label{sec:intro}

Energy input from active galactic nuclei (AGNs) is often invoked as a mechanism to regulate star formation in their host galaxies.  The discovery of X-ray cavities coincident with radio lobes in cool core galaxy clusters make a strong case for feedback limiting the growth of the most massive galaxies (\citealt{dutson14}; \citealt{hlavacek15}).  However, direct evidence of feedback is sparse among the $\sim90$\% of AGN that are radio-quiet \citep{ivezic02}. The growth of the central black holes found in almost all non-dwarf galaxies releases enough energy to disrupt star formation if at least $\sim$5\% of the AGN power output can be coupled to the galaxy interstellar medium (ISM) (\citealt{dimatteo05}; \citealt{hopkins06}).  \par
Searches for the physical mechanisms which can heat the ISM at this level are ongoing.  Powerful molecular outflows capable of suppressing star formation activity have been detected in Mrk 231 \citep{feruglio10} and NGC 1266 \citep{alatalo15} and a number of other AGNs \citep{fiore17}, but the driving mechanism of these outflows remains unclear and may differ in different galaxies.  \citet{feruglio10} find that the kinetic power of the outflow in Mrk 231 is a few percent of the AGN bolometric luminosity and consistent with originating from a highly supersonic shock produced by radiation pressure on the ISM.  \citet{alatalo15} suggest that the outflow in NGC 1266 is instead more likely to be driven by momentum coupling to the radio jet, and that star formation is suppressed by the injection of turbulence.  

\par
It has been suggested that AGN biconical outflows can impart $\sim5$\% of the AGN power into the ISM through termination shocks (\citealt{das06}; \citealt{fischer11}, hereafter \citetalias{fischer11}; \citealt{fischer13}, hereafter \citetalias{fischer13}; \citealt{crenshaw15}).  These biconical outflows, modeled from [OIII] line emission data, are randomly inclined with respect to the host galaxy and have large opening half angles of $\approx30-60^{\circ}$, with thicknesses of $\approx5-20^{\circ}$ \citepalias{fischer13}.  Such wide bicones are likely to intersect the host disk, interacting with the ISM.  All the outflows were modeled quite well with biconical models that are hollow along the central axes, exhibiting a pattern of linear acceleration to a maximum velocity, $v_{\mathrm{max}}$, and then a comparable or faster deceleration beginning at a turnover radius $r_t\approx0.1-1.1$ kpc \citepalias{fischer13}.  More recent models of these outflows have accounted for the rotational kinematics of the host galaxy disk, measuring mean maximum outflow radii of $0.6$ kpc as well as finding that the AGN can disturb the rotational kinematics of the gas out to mean distances of $1.1$ kpc (\citealt{fischer17}; \citealt{fischer18}).  In some cases, models have also allowed for outflowing material to be present along the central axis of the bicone when evidence suggests that the outflowing material originates in the galaxy disk (Mrk 573, \citealt{revalski18a}; Mrk 34, \citealt{revalski18b}).  The outflows can lose a large fraction of their kinetic energy (up to 75\%) over relatively small distances of $\approx20-200$ pc (\citetalias{fischer13}; \citealt{crenshaw15}; \citealt{revalski18a}; \citealt{revalski18b}; \citealt{revalski21}, hereafter \citetalias{revalski21}).  

\par
A plausible explanation for this deceleration is termination shocks due to interaction with the host interstellar medium (ISM), which convert the bulk kinetic energy of the outflow in turbulent motion and radiation.  Since no significant increase is observed in the [OIII] line widths beyond $r_t$, it is possible that most of the power is radiated away.  Given the typical outflow velocities of $v_{\mathrm{max}}\sim1000$ km s$^{-1}$, the expected temperature of the shocked emission is $kT\approx1.3(v_{\mathrm{shock}}/1000$ km s$^{-1})^2$ keV $\approx1.3$ keV \citep{raga02}, and thus should emit in the soft X-ray band.  

\par
Studies of the X-ray emission in NGC 4151 \citep{wang11b}, Mrk 573 \citep{paggi12}, NGC 1068 \cite{wang12}, and NGC 3393 \citep{maksym19} using \textit{Chandra} imaging find evidence of shocked X-ray emission near $r_t$ with a wide range of power input from shocks from about $0.05$\% of the AGN luminosity in Mrk 573 to $\sim$0.5\% in NGC 4151.  This level of AGN feedback from shocks may be sufficient to suppress star formation in the two-stage feedback model proposed by \citet{hopkins10}, but is significantly lower than the fraction of accretion power required by most AGN feedback models. 
\par

In this paper, we present a study of AGN feedback in \object{Mrk 78} using \textit{Chandra} imaging.  Of the 17 biconical outflows modeled by \citetalias{fischer13}, only three have high $v_{\mathrm{max}}>1000$ km s$^{-1}$, corresponding to high $kT$ temperature shocks easily detectable by \textit{Chandra}, and $r_t>0.5^{\prime\prime}$, which can be easily resolved by \textit{Chandra}.  One of these, NGC 1068, was previously observed by \textit{Chandra} and studied by \citet{wang12}.  The other sources are Mrk 78 (presented here) and Mrk 34 (to be presented in Maksym et al., in prep), for which the modeling by \citetalias{fischer11} and \citetalias{fischer13} indicate outflows in the plane of the host disk, so that ISM interactions are highly likely. Both AGN were targeted by a \textit{Chandra} program in 2017 (on January 1 and 7 for Mrk 78) in order to search for evidence of termination shocks at $r_t$ and study their impact on the ISM.  

\par
Mrk 78 is classified as an SB Seyfert 2 galaxy in the NASA/IPAC Extragalactic Database (NED).  It has a redshift of $z=0.03715$ \citep{michel88}, residing at a distance of $\approx160$ Mpc (for $H_0=70$ km s$^{-1}$ Mpc$^{-1}$).  There is evidence that Mrk 78 is heavily obscured based on its X-ray and mid-infrared (IR) properties.  Its WISE colors ($W1-W2>0.8$) indicate its mid-IR emission is dominated by AGN-heated dust \citep{stern12}, and its 12-micron luminosity ($\nu L_{\nu}$) is $L_{12}\approx9\times10^{43}$ erg s$^{-1}$.  It has been been shown to be nearly Compton-thick based on fitting the broadband X-ray spectrum using \textit{XMM} and \textit{NuSTAR} data \citep{zhao20}.  Its intrinsic $2-10$ keV luminosity is estimated to be $0.8-1.2\times10^{43}$ erg s$^{-1}$ \citep{zhao20}, making it consistent with the mid-IR to X-ray correlation from \citet{gandhi09}. \par
 At a distance of 160 Mpc, 1$^{\prime\prime}$ corresponds to a physical size scale of 735 pc.  The extended narrow-line region (NLR) in Mrk 78 and its relationship to the central radio source have been extensively studied (\citealt{whittle02}; \citet{whittle04}, hereafter \citetalias{whittle04}; \citealt{whittle05}, hereafter \citetalias{whittle05}; \citetalias{fischer11}).  \citetalias{fischer11} modeled the [OIII] line velocities measured from \textit{Hubble} Space Telescope Imaging Spectrograph (STIS) spectra using a biconical outflow model.  They find $v_{\mathrm{max}}=1200$ km s$^{-1}$ and $r_t=700$ parsecs.  A more recent analysis by \citet{revalski21} found a similar best-fit model for the [OIII] outflow, with a slightly larger turnover radius of $r_t=900$ parsecs.
\par
However, the outflow in Mrk 78 may be substantially more complex.  The [OIII] emission exhibits some high velocities and large line widths beyond the modeled $r_t$ on the Western side which are not well described by the outflow model.  Furthermore, there is significant asymmetry in both the [OIII] and radio emission on the East and West sides of the nucleus.  Based on the radio and [OIII] morphology as well as the [OIII] kinematics, \citetalias{whittle04} posited that on the Eastern side, the radio jet is deflected by the ionized gas it encounters and accelerates it.  In contrast, they suggest that on the Western side, near to or coincident with where \citetalias{fischer11} later found large velocities and line widths, the radio jet is disrupted by a compact cloud and expands into a leaky ``bubble", accelerating and ablating ionized gas knots until it blows out of the region through gaps in the gas.  
\par
In this paper, we use \textit{Chandra} imaging of Mrk 78 to search for evidence of shocks, and through comparison of the X-ray, radio, and [OIII] morphologies, better piece together the physical mechanisms by which the central AGN impacts the gas in the host galaxy.  In \S\ref{sec:obs}, we describe the \textit{Chandra} observations and pre-existing multi-wavelength data used in this work, as well as their astrometric registration.  The production of X-ray images is detailed in \S\ref{sec:imaging}, and the analysis of the X-ray morphology is discussed in \S\ref{sec:imagingresults}.  \S\ref{sec:spectral} describes the models and methods using for spectral analysis, and \S\ref{sec:spectralresults} provides the spectral results.  In \S\ref{sec:discussion}, we discuss the contribution of the \textit{Chandra} observations to our understanding of AGN feedback in Mrk 78, especially with regards to the presence and energetics of shocks.  

\section{Observations and Data Reduction}
\label{sec:obs}

\subsection{Chandra observations}

The \textit{Chandra X-ray Observatory} performed two observations of Mrk 78 in January 2017 with the Advanced CCD Imaging Spectrometer (ACIS).  The combined exposure of the two observations is 99.4 ks, and information about the individual observations is provided in Table \ref{tab:chandraobs}.  The observations were analyzed using CIAO 4.11 and CALDB 4.8.3.  
\par

\begin{table*}
\begin{minipage}{\textwidth}
\centering
\footnotesize
\caption{\textit{Chandra} Observations}
\begin{tabular}{ccccc} \hline \hline
\multirow{2}{*}{ObsID} & R.A. (J2000) & Decl. (J2000) & Start Time & Exposure \\
& (deg) & (deg) & (UT) & (ks) \\ \hline
18122 & 115.664589 & 65.173994 & 2017 Jan 7 03:50:14 & 49.85\\
19973 & 115.663627 & 65.173887 & 2017 Jan 1 21:13:50 & 49.59\\
\hline \hline
\end{tabular}
\label{tab:chandraobs}
\end{minipage}
\end{table*}

After running the \texttt{chandra\_repro} script to process the observations, we checked for background flares in each observation.  No significant flares that exceeded the average background rate by $>20\%$ were found.  Before combining together the two observations, we improved their relative astrometry.  To perform this astrometic correction, we first used the \texttt{fluximage} script to create $0.5-7$ keV images, exposure maps, and PSF maps of the observations and performed source detection using \texttt{wavdetect} in each image.  Then, we used \texttt{wcs\_match} to calculate the translational offsets between the two \textit{Chandra} observations by cross-matching the lists of approximately 20 sources detected with $>5\sigma$ significance, excluding Mrk 78.  We applied this astrometric adjustment using \texttt{wcs\_update}.  Finally, merged event files and a $0.3-3$ keV band image was produced using \texttt{merge\_obs}. 
\par
\bigskip
\bigskip
\subsection{[OIII] and 3.6 cm images}

In order to build up a more complete view of AGN feedback processes in Mrk 78, in this study we also make use of [OIII] and 3.6 cm images published in \citetalias{whittle04}.  Here we briefly describe how these images were produced, while full details are provided in \citetalias{whittle04}.  

\subsubsection{HST [OIII] image}

The [OIII] image in \citetalias{whittle04} was produced using observations from the pre-COSTAR Planetary Camera on the \textit{Hubble Space Telescope} (\textit{HST}). These observations were performed on August 29, 1992, with a total integration time of 5014 seconds in the F517N filter and 1800 seconds in F588N.  The F517N filter covers both [OIII] $\lambda5007$ \AA\ and 4949 \AA, while the F588N filter provides a continuum measurement.  The continuum was subtracted from the F517N image and the image was then deconvolved.  \citetalias{whittle04} found that the continuum and line flux measurements based on their images were in excellent agreement with ground-based measurements from \citet{debruyn78}.  

\subsubsection{VLA 3.6 cm image}

The 3.6 cm image from \citetalias{whittle04} was produced from observations by the Karl G. Jansky Very Large Array (VLA) taken in 1990 with A array with a total integration time of 8 hours.  The data was flux calibrated.  In this work, we use the map that \citetalias{whittle04} produced using natural weighting, yielding a beam of $0.29^{\prime\prime}\times0.27^{\prime\prime}$ in P.A.$=12^{\circ}$ and a noise level of 9 $\mu$Jy beam$^{-1}$.

\subsection{Astrometric registration}

In order to be able to compare the positions of features in the \textit{Chandra}, \textit{HST}, and VLA images, it is important for the astrometry of each image to be as accurate as possible.  To improve the absolute astrometry of the \textit{Chandra} observations, we re-ran the CIAO tool \texttt{wavdetect} on the merged $0.3-3$ keV image and searched for multi-wavelength counterparts to all \textit{Chandra} sources detected with $>5\sigma$ significance using Vizier.  We found unique counterparts within 1$^{\prime\prime}$ for four of the \textit{Chandra} sources, which are listed in Table \ref{tab:counterparts}.  We used these counterparts to update \textit{Chandra}'s astrometry using \texttt{wcs\_match} and \texttt{wcs\_update}.  The average residual offset between the \textit{Chandra} and optical/infrared counterparts decreased from $0\farcs52$ to $0\farcs16$ after this astrometric correction.  This average residual offset provides an estimate of the systematic astrometric uncertainty.  The peak of the $2-8$ keV emission is located at R.A. = 7:42:41.70, Dec = +65:10:37.46 (J2000); the statistical uncertainty associated with this position is $\pm0\farcs07$.  

\par
\begin{figure}
\centering
\includegraphics[width=0.45\textwidth]{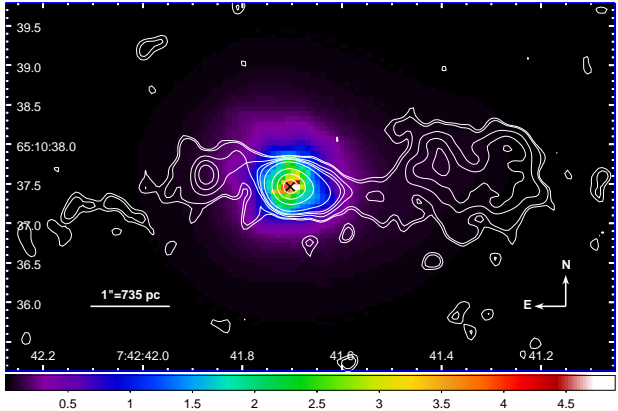}
\caption{\textit{Chandra} $2-8$ keV image binned by 1/8 subpixel and adaptively smoothed.  VLA 3.6cm contours are shown in white.  The radio core has been aligned with the $2-8$ keV peak shown by the black cross.}
\label{fig:hardxray}
\end{figure}

The VLA position of the 3.6 cm radio core is $0\farcs10$ N and $0\farcs08$ E of the peak of the $2-8$ keV emission. The VLA positional errors are $\pm0\farcs10$, so this position is consistent with the $2-8$ keV peak.  Since both the radio core and hard X-ray peak are expected to emanate from the vicinity of the AGN, we shifted the 3.6 cm image so that the radio core and hard X-ray peak coincide, as shown in Figure \ref{fig:hardxray}.  However, this adjustment is small enough that it does not significantly impact our results.

\par
For the \textit{HST} image, we adopt the same astrometry as \citetalias{whittle04}, who assumed no astrometric offset between the F517N and F588N images and fixed the \textit{HST} F588N continuum peak to the ground-based blue-green continuum peak \citep{clements81}.  The optical continuum peak is located at a distance of $0\farcs13$ from the 2-8 keV peak, which is consistent within the errors.  Since Mrk 78 has a prominent dust lane, seen in the blue F342W FOC image shown in \citetalias{whittle04}, it is possible for the hard X-ray peak and the optical continuum peak not to coincide due to obscuration.  Therefore, we do not adjust the astrometry of the optical images to force these two peaks to coincide.

\begin{table*}
\begin{minipage}{\textwidth}
\centering
\footnotesize
\caption{Optical and infrared counterparts used for astrometric registration}
\begin{tabular}{ccccc} \hline \hline
\multirow{2}{*}{Catalog} & R.A. (J2000) & Decl. (J2000) & Initial offset from \textit{Chandra} & Magnitudes \\
& (deg) & (deg) & (arcsec) & (Vega) \\ \hline

AllWISE & 115.6634718 & 65.1474536 & 0.70 & W1=$16.41\pm0.06$, W2=$15.7\pm0.1$, W3=$12.2\pm0.3$\\
Gaia & 115.753066 & 65.185329 & 0.56 & G=$20.67\pm0.01$, BP=$20.73\pm0.08$, RP=$19.48\pm0.05$\\
Gaia & 115.819352 & 65.158297 & 0.25 & G=$19.734\pm0.006$, BP=$19.85\pm0.04$, RP=$19.29\pm0.04$\\
USNO-B1.0 & 115.631550 & 65.215975 & 0.57 & B1=20.46, R1=18.85, B2=19.68, R2=18.56\\
\hline \hline
\multicolumn{5}{p{7.0in}}{\T Notes: 
References for catalogs: AllWISE \citep{cutri13}, Gaia DR2 (Gaia Collaboration 2018), USNO-B1.0 (Monet+2003)
} \\
\end{tabular}
\label{tab:counterparts}
\end{minipage}
\end{table*}

\begin{figure}
\centering
\includegraphics[width=0.45\textwidth]{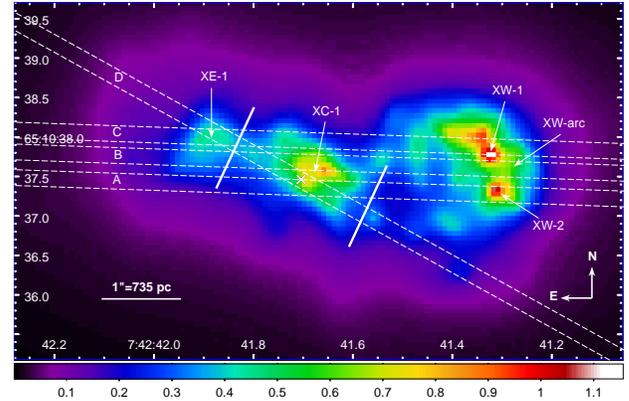}
\caption{\textit{Chandra} $0.3-2$ keV image binned by 1/8 subpixel and adaptively smoothed.  The location of the $2-8$ keV peak is shown by a white cross. The white solid lines indicate the turnover radius of the [OIII] outflow model from \citetalias{fischer11}, and the white dashed lines show the STIS slit locations and slit names as used by \citetalias{fischer11}.  Soft X-ray features are labeled to facilitate discussion in the text. }
\label{fig:xraylabels}
\end{figure}

\bigskip
\bigskip
\bigskip
\section{Imaging Analysis}
\label{sec:imaging}

After astrometrically registering the \textit{Chandra} observations, we produced images in the $0.3-2$ keV and $2-8$ keV bands with 1/8 pixel scale to study the detailed structure of the emission at the center of Mrk 78.  We adaptively smoothed the images with the \texttt{ciaoadapt} tool in ds9, using Gaussian smoothing kernels with radii ranging from $0.5-15$ pixels.  Comparing the results using different values of the minimum counts under each kernel, we find that a value of 11 counts reveals the highest resolution features that consistently appear in smoothed images.  The resulting $0.3-2$ keV and $2-8$ keV smoothed images are shown in Figures \ref{fig:hardxray}-\ref{fig:multiwavelength}.  Figure \ref{fig:xraylabels} includes labels for prominent soft X-ray features.  

\par
We explore how the morphology of the soft X-ray emission varies with energy.  We produced adaptively smoothed images in the $0.3-1$ keV and $1-2$ keV bands, in the same manner as described above but with a lower minimum count per kernel value of 10.  These images are included in the three-color X-ray image shown in Figure \ref{fig:xray3color}.  In order to search for possible evidence of shocked emission, we also made a map of the $0.82-0.92$ keV band, the redshifted energy of the Ne IX line (rest-frame=0.905 keV), which is often observed in shocked emission regions, correlated with radio jets in other AGN (e.g., NGC4151, \citealt{wang11a,wang11b,wang11c}; Mrk 573, \citealt{paggi12}; NGC 3393, \citealt{maksym19}).  This smoothed map, shown in Figure \ref{fig:neiximages}, is made in the same manner as the other images, except that the minimum counts per kernel value was lowered to 5.  

\par
In addition, since the $L$[OIII]/$L_{\mathrm{X}}$ ratio can help distinguish between photionized and shock-ionized emission (\citet{bianchi06}, \citep{wang12}), we also made a map of the $L$[OIII]/$L_{\mathrm{X}}$ ratio.  This map was made by binning both the [OIII] and $0.5-2$ keV images by the same grid of $0\farcs246$ pixels using the \textit{CIAO} tool \texttt{reproject\_image\_grid}.  The $0.5-2$ keV counts image was converted into a flux image using a conversion factor of $8.35\times10^{-17}$ erg s$^{-1}$ cm$^{2}$ photon$^{-1}$ based on spectral fitting of the total $0.5-2$ keV emission of Mrk 78 (see \S\ref{sec:spectral}).  Then using \texttt{dmimgcalc}, we divided the [OIII] flux image by the $0.5-2$ keV flux image to obtain the $L$[OIII]/$L_{\mathrm{X}}$ ratio map shown in Figure \ref{fig:oiiixraymap}.

\section{Imaging Results}
\label{sec:imagingresults}

\subsection{X-ray morphology}
As can seen in Figure \ref{fig:xraylabels}, the $0.3-2$ keV emission at the center of Mrk 78 is extended, especially in the E-W direction.  There is a bright knot of $0.3-2$ keV emission labeled XC-1 in Figure \ref{fig:xraylabels} that is roughly 1$^{\prime\prime}$ across in the vicinity of the nucleus.  As expected given that Mrk 78 hosts an obscured Seyfert 2 AGN, this emission does not originate directly from the innermost part of the AGN (i.e. accretion disk or corona) as evidenced by the fact that it is not centered on the peak of the $2-8$ keV emission and it has a more asymmetric profile than expected for the \textit{Chandra} PSF. The morphology of the extended emission more than $1^{\prime\prime}$ away from the Mrk 78 nucleus is very different on the Eastern and Western sides.  On the Eastern side, the emission is fairly compact, peaking in knot XE-1 located about $1\farcs3$ (950 pc) from the nucleus.  On the Western side, the emission spreads out into a wider feature, extending about $3\farcs0$ (2200 pc) from the nucleus and exhibiting a curved arc of X-ray emission about $1^{\prime\prime}$ in length on the outer edge, with two particularly bright X-ray knots, labeled XW-1 and XW-2 in Figure \ref{fig:xraylabels}.  
\par
The observed X-ray morphology is most likely associated with interactions between the AGN and the surrounding medium.  It is unlikely that the X-ray emission is associated with star formation as the host galaxy of Mrk 78 exhibits optical spectral signatures indicative of being in a post-starburst phase with stellar populations $>10^8$ years old \citep{fernandes01}. The X-ray morphology of Mrk 78 is also unlikely to be affected by external factors such as ongoing mergers given that no low surface brightness extended features have been detected \citep{smirnova10}, and that the nearest neighboring galaxy appearing in R-band images of Mrk 78 has a much higher spectroscopic redshift of $z=0.38$ \citep{kozlova20}. 

\begin{figure}
\centering
\includegraphics[width=0.45\textwidth]{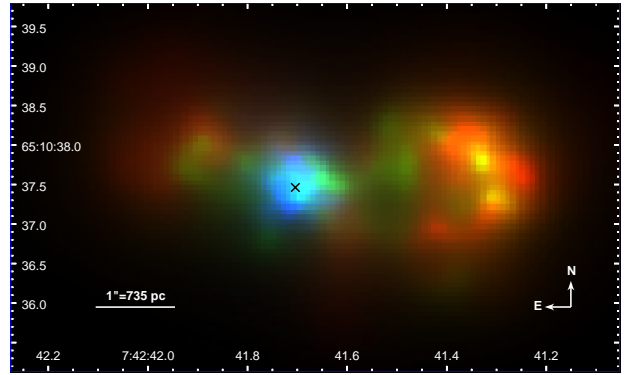}
\caption{Adaptively smoothed three-color X-ray image showing the $0.3-1$ keV band in red, $1-2$ keV band in green, and $2-8$ keV band in blue.  The cross represents the peak of the $2-8$ keV emission. All images are linearly scaled between a minimum value of zero counts and a maximum value of 0.3, 0.6, and 2.0 counts per pixel for the $0.3-1$, $1-2$, and $2-8$ keV bands.}
\label{fig:xray3color}
\end{figure}

\begin{figure*}
\gridline{\fig{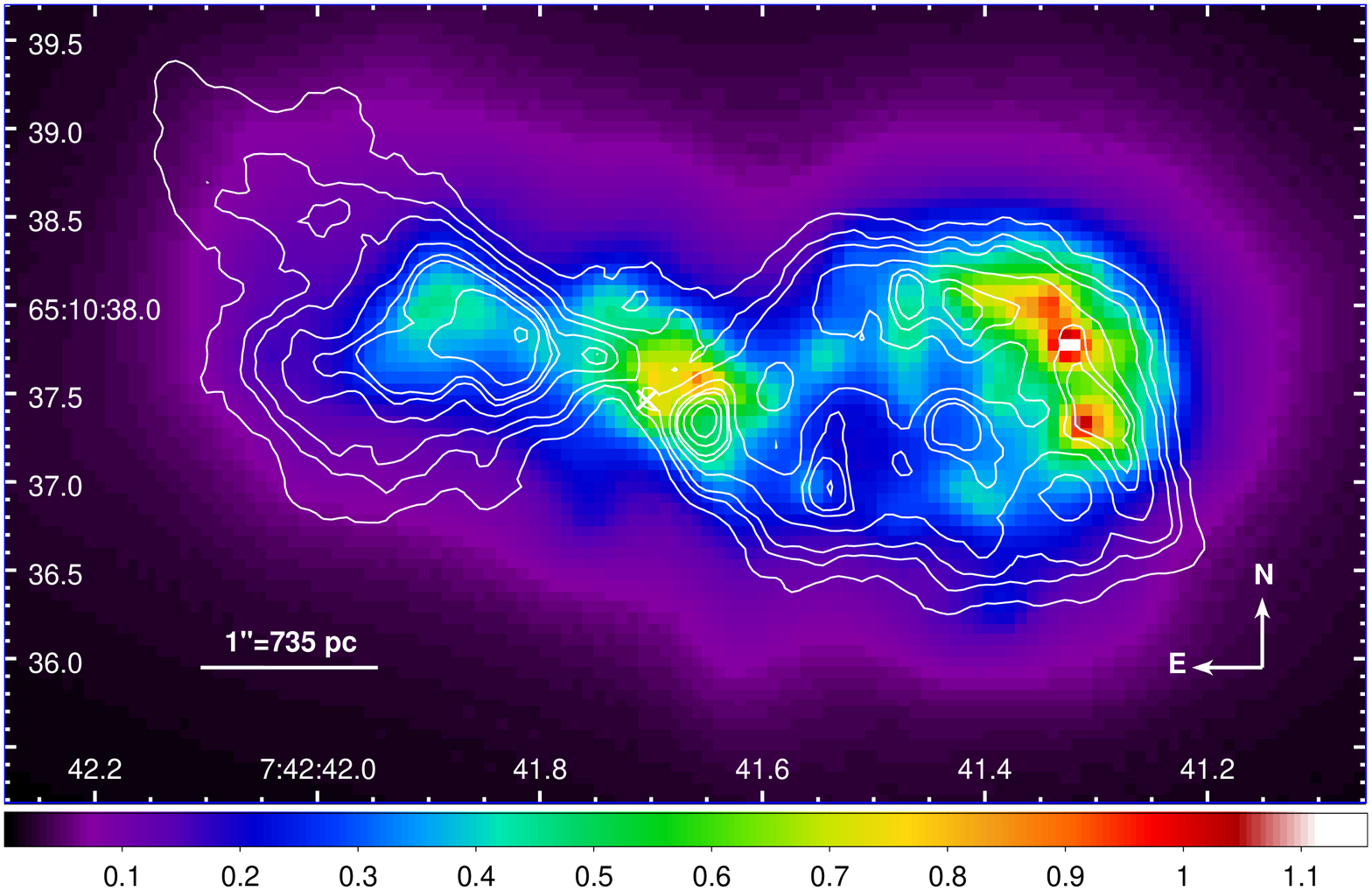}{0.45\textwidth}{(a) }
          \fig{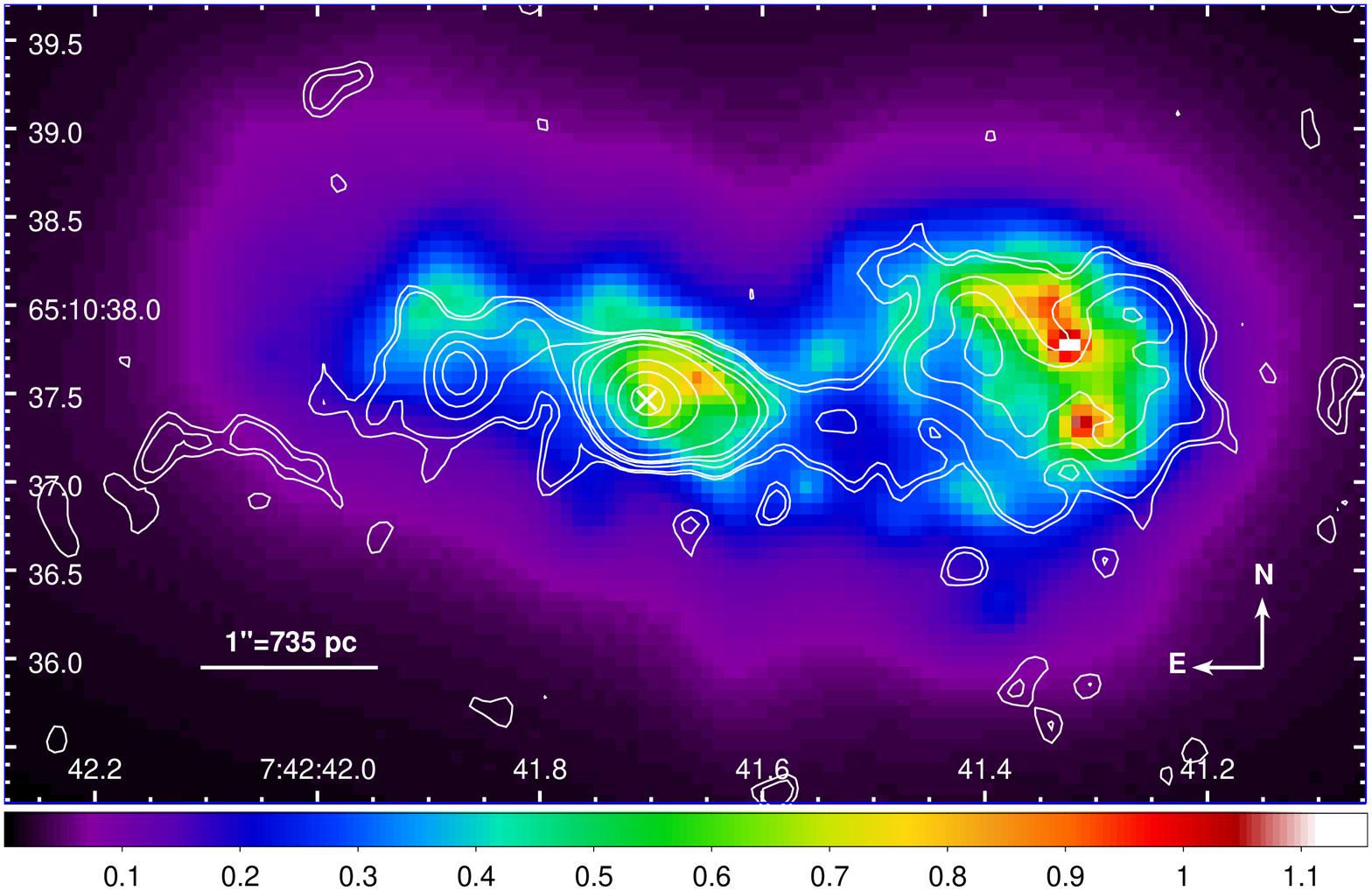}{0.45\textwidth}{(b) }
          }
\caption{Both panels display the $0.3-2$ keV smoothed counts image with 1/8 subpixel binning.  In panel (a), the white contours show the [OIII] emission, and the cross represents the peak of the $2-8$ keV emission.  In panel (b), the white contours show the 3.6 cm emission.
\label{fig:multiwavelength}}
\end{figure*}

\subsection{X-ray morphological variation with energy}
The X-ray morphology of the extended emission varies with energy.  As shown in Figure \ref{fig:xray3color}, the the extended emission located $\lesssim2^{\prime\prime}$ from the nucleus is harder, having a higher ratio of $1-2$ keV flux to $0.3-1$ keV flux, compared to the outer regions of the extended emission. The two X-ray knots on the Western edge, XW-1 and XW-2, also exhibit harder X-ray emission, having a higher ratio of $1-2$ keV flux to $0.3-1$ keV flux than the rest of the arc emission. 
\par
\begin{figure*}
\gridline{\fig{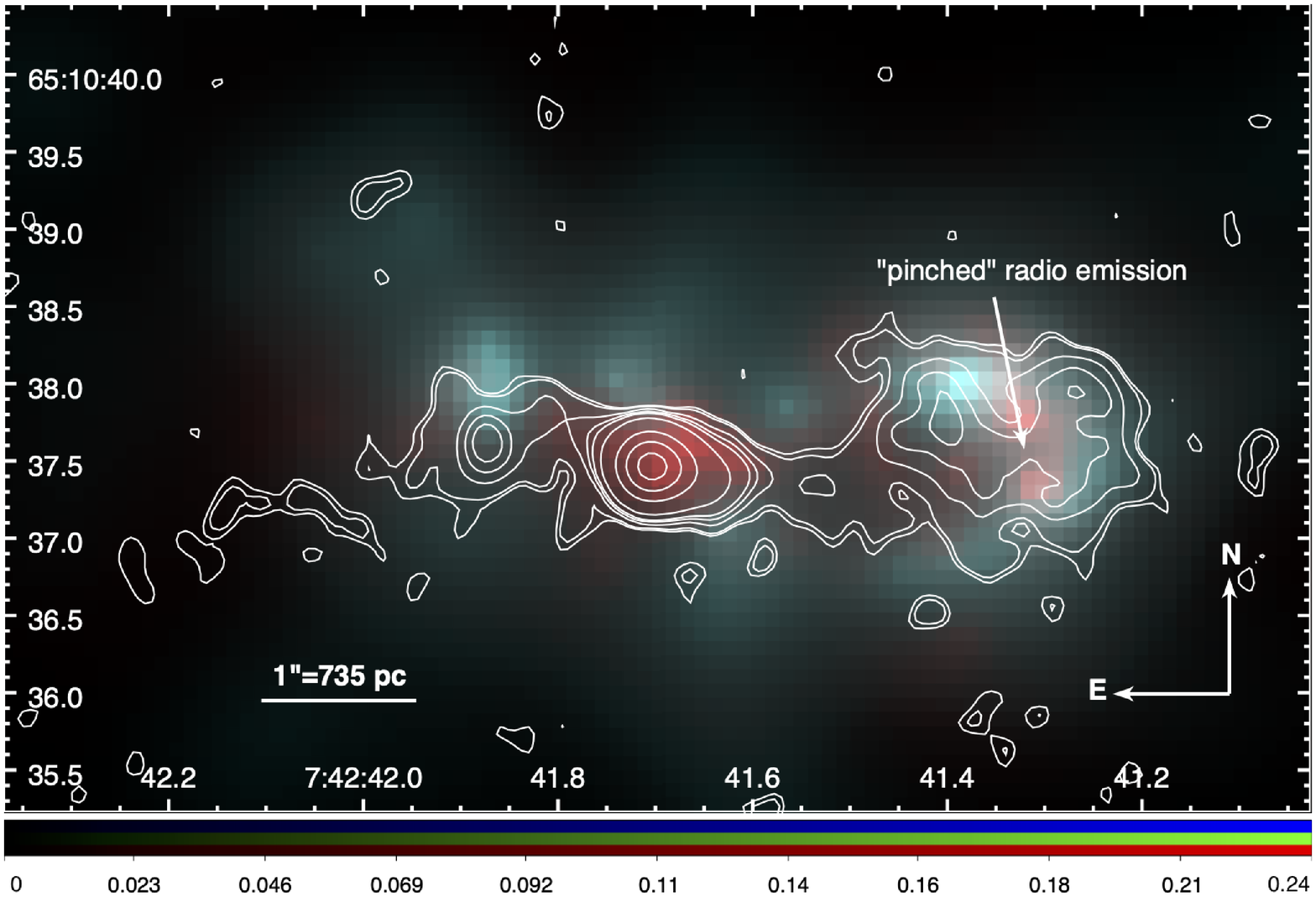}{0.45\textwidth}{(a)}
          \fig{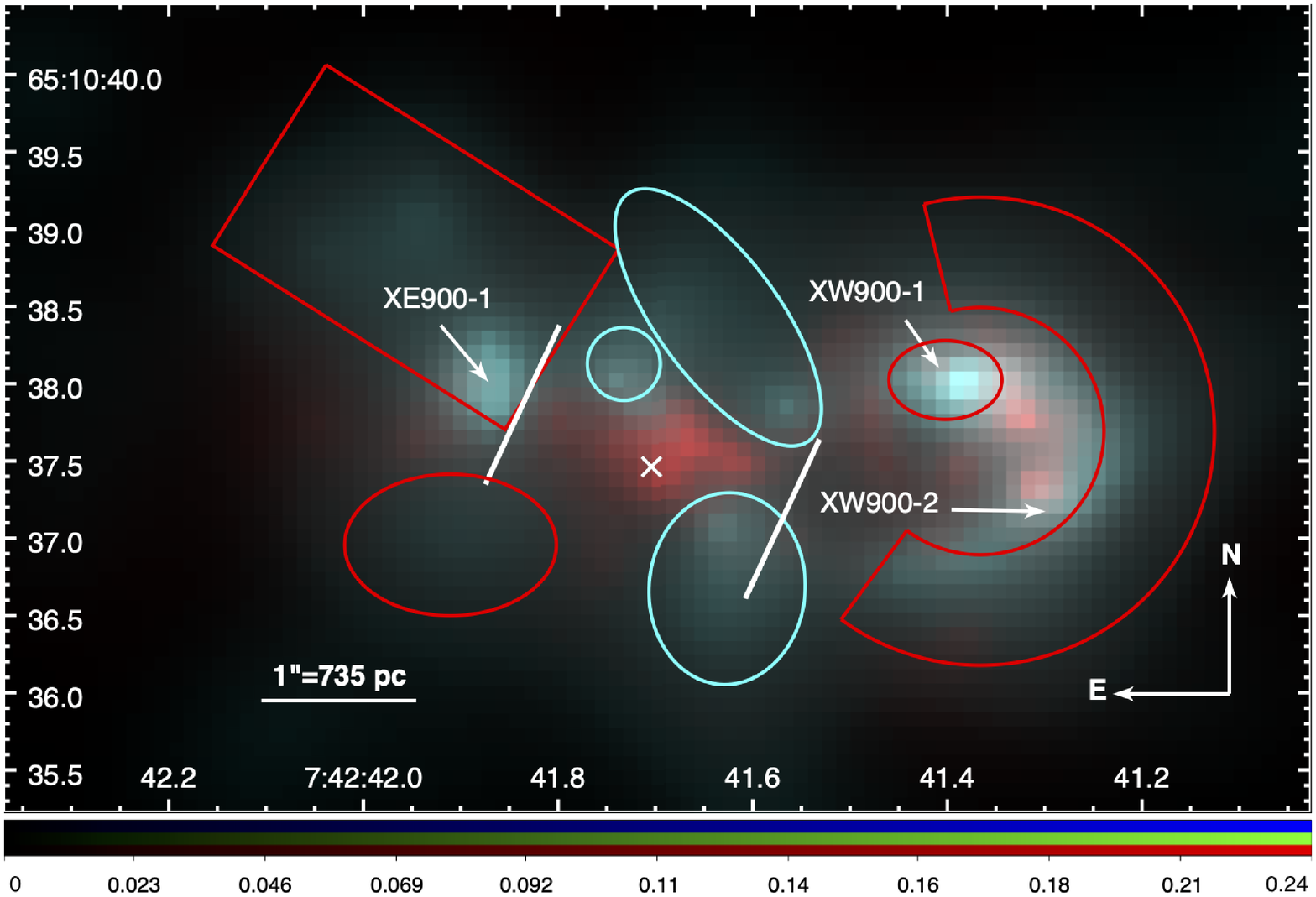}{0.45\textwidth}{(b)}
          }
\caption{Both panels show smoothed counts images of 900 eV emission within $\pm50$ eV of the NeIX line energy in green/blue and the remainder of the $0.3-2$ keV emission in red.  In panel (a), the white contours show the 3 cm contours.  In panel (b), the cross represents the peak of the $2-8$ keV emission, the white lines indicate the turnover radius of the [OIII] outflow model from \citetalias{fischer11}, the cyan regions show the 900 eV spectral extraction regions near the nucleus and the red regions indicate the 900 eV spectral extraction regions farther out from the nucleus.  The colorbar values shown correspond to the red color image; the scale for the green/blue image is approximately a factor of 7.5 lower than for the red image.
\label{fig:neiximages}}
\end{figure*}

The emission within $\pm50$ eV of the rest-frame Ne IX energy (905 eV) shown in Figure \ref{fig:neiximages} exhibits different morphology than the rest of the $0.3-2$ keV emission shown by the contours.  We will refer to this emission in the observed $822-922$ eV band as 900 eV emission.  Within $1^{\prime\prime}$ of the nucleus, the 900 eV emission exhibits three faint knots which are offset from the central $0.3-2$ keV peak.  On the Eastern side, there is a knot of 900 eV emission (XE900-1) roughly coincident with XE-1, which is close to the [OIII] velocity turnover radius from the \citetalias{fischer11} and \citetalias{revalski21} outflow models; faint 900 eV emission appears to extend farther to the NE than the bulk of the $0.3-2$ keV emission.  On the Western side, the 900 eV emission displays a similar arc as the $0.3-2$ keV emission, but the brightest 900 eV knot XW900-1 is not coincident with either of the two knots (XW-1 and XW-2) within the arc.  Thus, some regions of extended emission may exhibit enhanced Ne IX emission, which is indicative of shocks.  \par


\begin{figure*}
\centering
\includegraphics[width=1.0\textwidth]{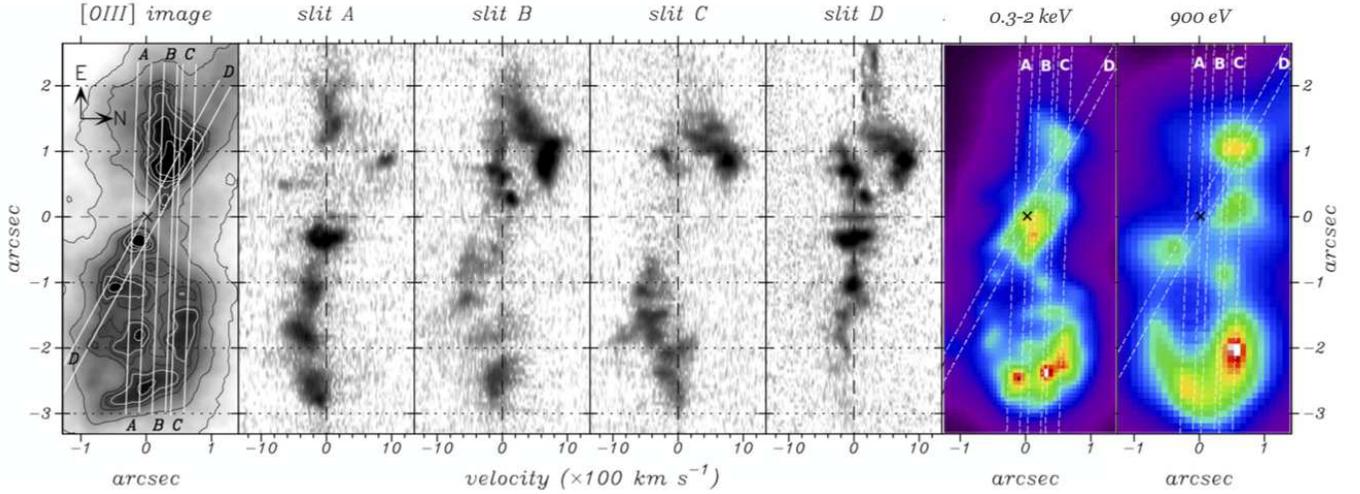}
\caption{From left to right, images of the [OIII] emission, the [OIII] velocity field in the four STIS slits, the $0.3-2$ emission, and the 900 eV emission.  The positions of the STIS slits are overlayed on the [OIII] and X-ray images.  The color scales of the X-ray images are chosen so that key features are easily visible; the colorbar values of the 900 eV emission are a factor of 10 lower than those of the $0.3-2$ keV emission.}
\label{fig:oiiivel}
\end{figure*}
\subsection{Comparison of X-ray, [OIII], and radio morphology}
\label{sec:morphcomparison}
As shown in Figure \ref{fig:multiwavelength}, the E-W asymmetry of the soft X-ray emission resembles the asymmetry seen in the [OIII] and 3.6 cm maps.  On the Eastern side, the X-ray, [OIII], and radio knots are in close proximity to one another; the peak of the X-ray emission (XE-1) lies farther from the nucleus than the peak of the [OIII] emission.  However, the X-ray emission does not extend as far to the NE as the [OIII] emission, but is truncated at roughly the location where the radio emission begins to bend towards the SE. \par 
On the Western side, both the [OIII] and radio maps show a ``bubble"-like morphology similar to the soft X-rays.  However, in detail their morphology differs.  There are several significant knots of [OIII] emission between $1-2^{\prime\prime}$ from the nucleus, where both the soft X-ray and radio emission is relatively low.  Both the X-ray and [OIII] emission exhibit a bright, curved arc $2-3^{\prime\prime}$ West of the nucleus, however the two brightest X-ray knots within this region (XW-1 and XW-2) are offset from the [OIII] peaks within this arc.  The radio emission is bright on the inside and outside of this curved arc and in between the two brightest X-ray knots within the arc.  These two X-ray knots coincide with areas where the radio emission appears ``pinched" towards the center of the E-W radio axis.  The 900 eV peaks (XW900-1, XW900-2) in the Western arc also roughly coincide with bright [OIII] emission in the Western arc and the location where the Western radio emission is pinched, as can be seen in Figure \ref{fig:neiximages}.  \par
The white lines in Figure \ref{fig:xraylabels} show the location of the turnover radius from the modeling of the [OIII] outflow by \citetalias{fischer11} and \citetalias{revalski21}.  This is the location at which the outflow model reaches maximum velocity and then begins to decelerate.  Near this radius, there is a knot of X-ray emission on the Eastern side of the nucleus, but an overall dearth of X-ray emission on the Western side.  This radius appears to coincide with changes in the radio morphology, with the bending of the radio jet on the Eastern side and the beginning of the jet's expansion into a wider structure on the Western side. 
\par
Figure \ref{fig:oiiivel} compares the [OIII] velocity field in each of the four \textit{HST} STIS slits, the $0.3-2$ keV emission, and the 900 eV emission.  As can be seen, the large [OIII] velocity decrease seen $1^{\prime\prime}$ East of the nucleus in all the slit velocity fields coincides with the location of the Eastern knot XE-1, visible both in $0.3-2$ keV and 900 eV emission maps.  Another large [OIII] velocity drop is visible $2^{\prime\prime}$ West of the nucleus in slit C, which coincides with the brightest 900 eV knot XW900-1.  This velocity drop is not accounted for by the \citetalias{fischer11} or \citetalias{revalski21} [OIII] biconical outflow models, indicating that this model may be an incomplete description of the [OIII] outflow kinematics and that Mrk 78 may require more complex modeling as has been more recently performed for other AGN (\citealt{fischer17};\citealt{fischer18}; \citealt{revalski18a}; \citealt{revalski18b}).

\subsection{The $L$[OIII]/$L_{\mathrm{X}}$ ratio}
\begin{figure}
\centering
\includegraphics[width=0.45\textwidth]{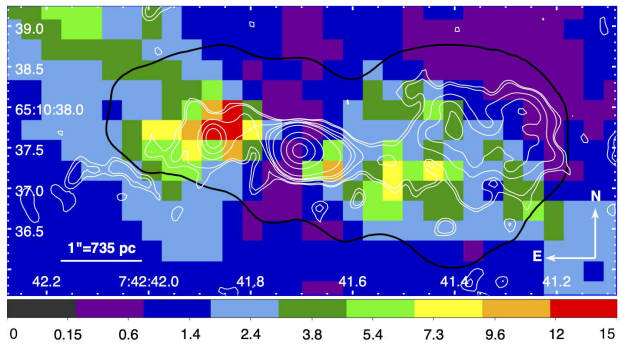}
\caption{Map of $L$[OIII]/$L_{\mathrm{X}}$ ratio, where $L_{\mathrm{X}}$ is measured in the $0.5-2$ keV band.  The white contours display the 3.6 cm emission contours.  The black lines denotes the region outside of which both the $0.3-2$ keV and [OIII] emission is low and thus the ratio values are not significant.}
\label{fig:oiiixraymap}
\end{figure}

As shown in Figure \ref{fig:oiiixraymap}, the extended emission of Mrk 78 shows a range of $L$[OIII]/$L_{\mathrm{X}}$ values, with values generally decreasing outwards along the cross-cone direction.  \citet{bianchi06} modeled the dependence of the $L$[OIII]/$L_{\mathrm{X}}$ ratio for photoionized emission on the ionization parameter, $U=Q_{\mathrm{ion}}/(4\pi r^2 c n_e)$, where $Q_{\mathrm{ion}}$ is the number of hydrogen-ionizing photons emitted by the central object per second and $r$ is the distance to the central ionizing source.  If $U$ decreases with increasing radius from the source, $L$[OIII]/$L_{\mathrm{X}}$ is expected to increase, whereas if $U$ increases with radius, $L$[OIII]/$L_{\mathrm{X}}$ is expected to decrease.  As discussed in \S\ref{sec:spectralext}, our spectral analysis reveals that $U$ is either roughly constant or decreases with distance from the AGN.  Therefore, the lower values of $L$[OIII]/$L_{\mathrm{X}}$ in the Western arc cannot be explained by trends in the photoionization parameter. \par
One factor which can result in lower $L$[OIII]/$L_{\mathrm{X}}$ ratios than expected from photoionization is the presence of shocks (\citealt{wang12}; \citealt{maksym19}).  As discussed in \S\ref{sec:shocklocation}, shocks may be responsible for the low $L$[OIII]/$L_{\mathrm{X}}$ in the Western arc.  Another factor which can result in lower $L$[OIII]/$L_{\mathrm{X}}$ ratios is obscuration; the low $L$[OIII]/$L_{\mathrm{X}}$ ratios in the cross-cone regions North and South of the nucleus are consistent with the presence of a dust lane visible in optical images which obscures the optical emission more than the soft X-ray emission.

\smallskip

\bigskip
\section{Spectral Analysis}
\label{sec:spectral}

\begin{figure}
    \centering
    \includegraphics[width=0.45\textwidth]{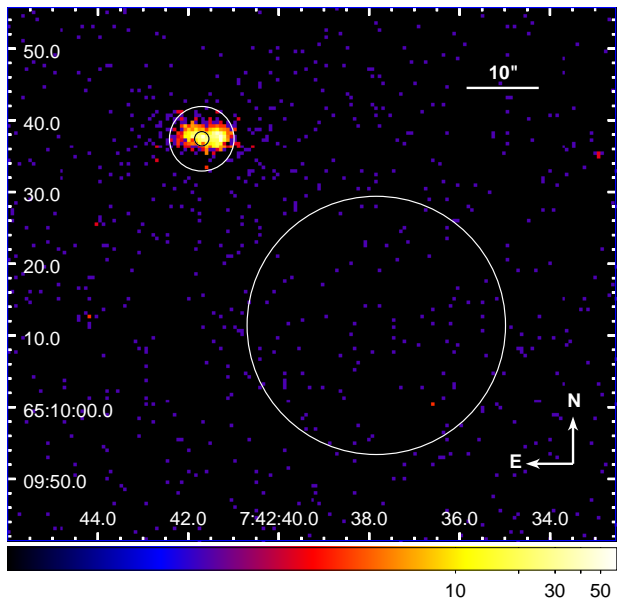}
    \caption{0.3-2 keV raw counts image.  The large $18^{\prime\prime}$-radius circle shows the background extraction region using for all our spectra.  A $1^{\prime\prime}$-radius circle shows the source region for the Mrk 78 nucleus, while the annulus with outer $4\farcs5$-radius shows the source region encompassing all the extended emission.}
    \label{fig:largespectralregions}
\end{figure}

\begin{figure}
    \centering
    \includegraphics[width=0.45\textwidth]{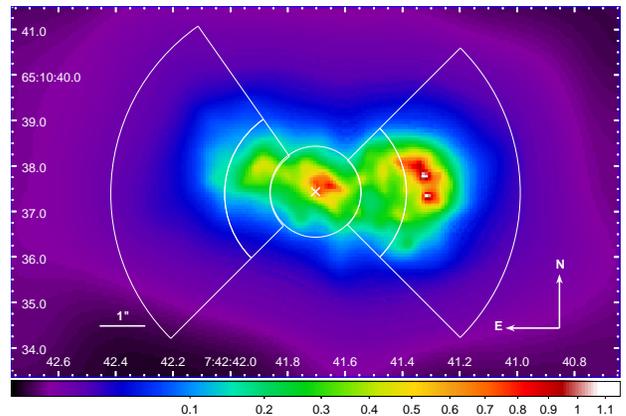}
    \caption{0.3-2 keV smoothed image with spectral extraction regions shown in white.  The cross shows the peak of the $2-8$ keV emission. The spectrum of the Mrk 78 nucleus is extracted from the circular region.  We extract spectra separately from the inner and outer regions of extended emission using the pie sector slices shown.}
    \label{fig:piespectralregions}
\end{figure}

We analyzed the spectrum of the soft X-ray emission to investigate its origin, in particular whether it arises from photoionization by the central AGN or collisional ionization, possibly associated with shocks.  Given the spatial variation of the X-ray properties discussed in \S\ref{sec:imaging}, we analyzed and compared the X-ray spectrum of different emission regions.  Figure \ref{fig:largespectralregions} shows the $18^{\prime\prime}$-radius circular background extraction region we used for all the spectra, which lies on the same ACIS-S chip as Mrk 78 and avoids any sources detected by \texttt{wavdetect}.  This figure also displays the $1^{\prime\prime}$-radius circular region used to extract the spectrum of the Mrk 78 nucleus and the $1^{\prime\prime}$ - $4\farcs5$-radius annular region encompassing the total extended soft X-ray emission.  As shown in Figure \ref{fig:piespectralregions}, we further split up this annular region into Eastern and Western pie sectors along the bi-cone axis; these sectors are divided into inner and outer regions that are $1-2^{\prime\prime}$ and $2 - 4.5^{\prime\prime}$ from the nucleus, respectively.  The inner regions have higher $L$[OIII]/$L_{\mathrm{X}}$ ratios and higher fractions of $1 - 2$ keV flux compared to $0.3 - 1$ keV flux, appearing green in Figure \ref{fig:xray3color}. \par
In addition, we also independently analyzed the spectra of regions with enhanced 900 eV emission.  Based on Figure \ref{fig:neiximages}, we identified spectral regions that exhibit elevated 900 eV emission compared to the bulk of the $0.3-2$ keV emission; these are the regions that appear cyan in Figure \ref{fig:neiximages}.  We selected as many of these regions as possible in order to obtain as many counts as possible for spectral analysis.  We analyzed regions within $1^{\prime\prime}$ of the nucleus (shown by cyan shapes) separately from those farther out from the nucleus (shown by red shapes), since we found that the $0.3-2$ keV emission differs significantly in these regions as discussed in \S\ref{sec:spectral}.\par
We extract the spectrum of each source region and corresponding ARF and RMF files using \texttt{specextract}.  We bin each spectrum so that each energy bin has at least 15 counts, and use chi-squared statistics when fitting the spectra.  We find that fitting the spectra binned by a minimum 5 counts per bin using the L-statistic (``lstat'' in XSPEC \citep{arnaud96}), which is appropriate for Poissonian data with Poissonian background, produces consistent results.
\par
In our spectral analysis of the soft X-ray emission, we use the XSPEC \texttt{apec} model to represent the thermal continuum and line emission from collisionally-ionized diffuse gas.  We freeze the relative metal abundances for these thermal models to solar values from \citet{anders89}.  To model photoionized gas emission we used an XSPEC grid of models produced using the CLOUDY c08.01 package \citep{ferland98}.  This model grid is the same as that used by \citet{paggi12} and \citet{maksym19} in studies of the soft X-ray emission associated with AGN in Mrk 573 and NGC 3393.  The ionization source for this CLOUDY grid was assumed to be an AGN continuum with a ``big bump'' temperature of $10^6$ K, an X-ray to UV ratio $\alpha_{OX}=-1.30$, and an X-ray powerlaw component with spectral index $\alpha=-0.8$.  The emitting cloud was assumed to have plane-parallel geometry and constant electron density $n_e=10^5$ cm$^{-3}$.  Note that the fraction of ionized species and equilibrium populations of excited states for key elements are expected to be similar for electron densities $n_e\sim1-10^6$ cm$^{-3}$ (\citealt{porquet00}; \citealt{ferland17}).  The model grid was parametrized in terms of the ionization parameter ($U=Q_{\mathrm{ion}}/(4\pi r^2 c n_e)$) over the range log$U$=[-3.0:2.0] in steps of 0.25 and the hydrogen column density ($N_{\mathrm{H}}$) over the range log$N_{\mathrm{H}}$=[19.0:23.5] in steps of 0.1.  
\par
For each spectrum, we begin by fitting a single component model subject only to Galactic absorption, which we estimate using the COLDEN tool\footnote{https://cxc.harvard.edu/toolkit/colden.jsp} to be $N_{\mathrm{H,Gal}}=4.1\times10^{20}$ cm$^{-2}$.  More components were added, one at a time if: (1) the best fit resulted in a reduced chi-squared value $\chi_{\nu}^2>2$, (2) a null hypothesis probability, $p_{\mathrm{null}} < 5$\%, or (3) there was significant structure in the residuals. This process was continued until a good fit was produced.

\section{Spectral Results}
\label{sec:spectralresults}
\subsection{Mrk 78 Nucleus}

\begin{figure*}
\gridline{\hspace{-1in}\rotatefig{270}{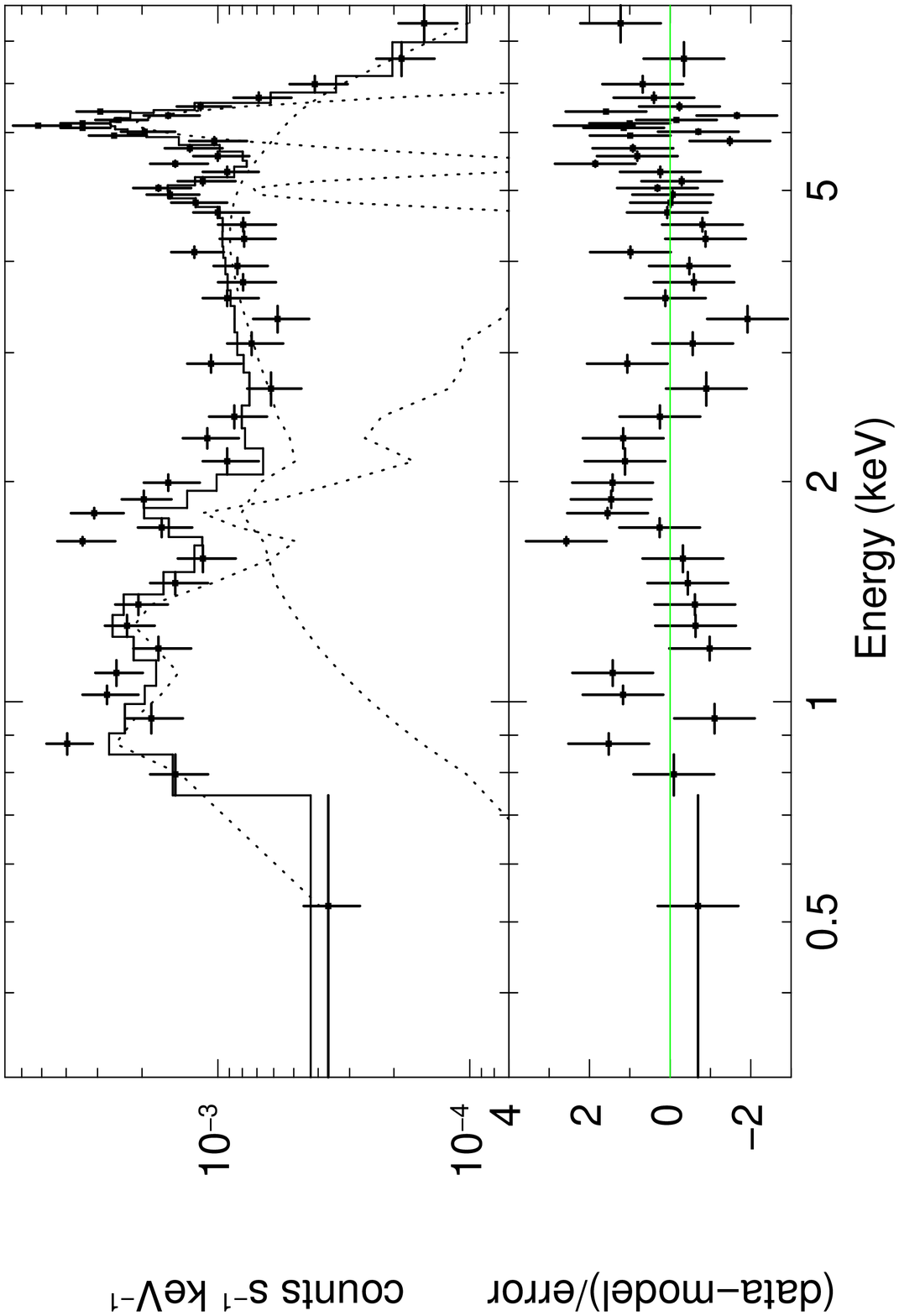}{0.2\textwidth}{\hspace{1in}(a)}
          \hspace{-1in}\rotatefig{270}{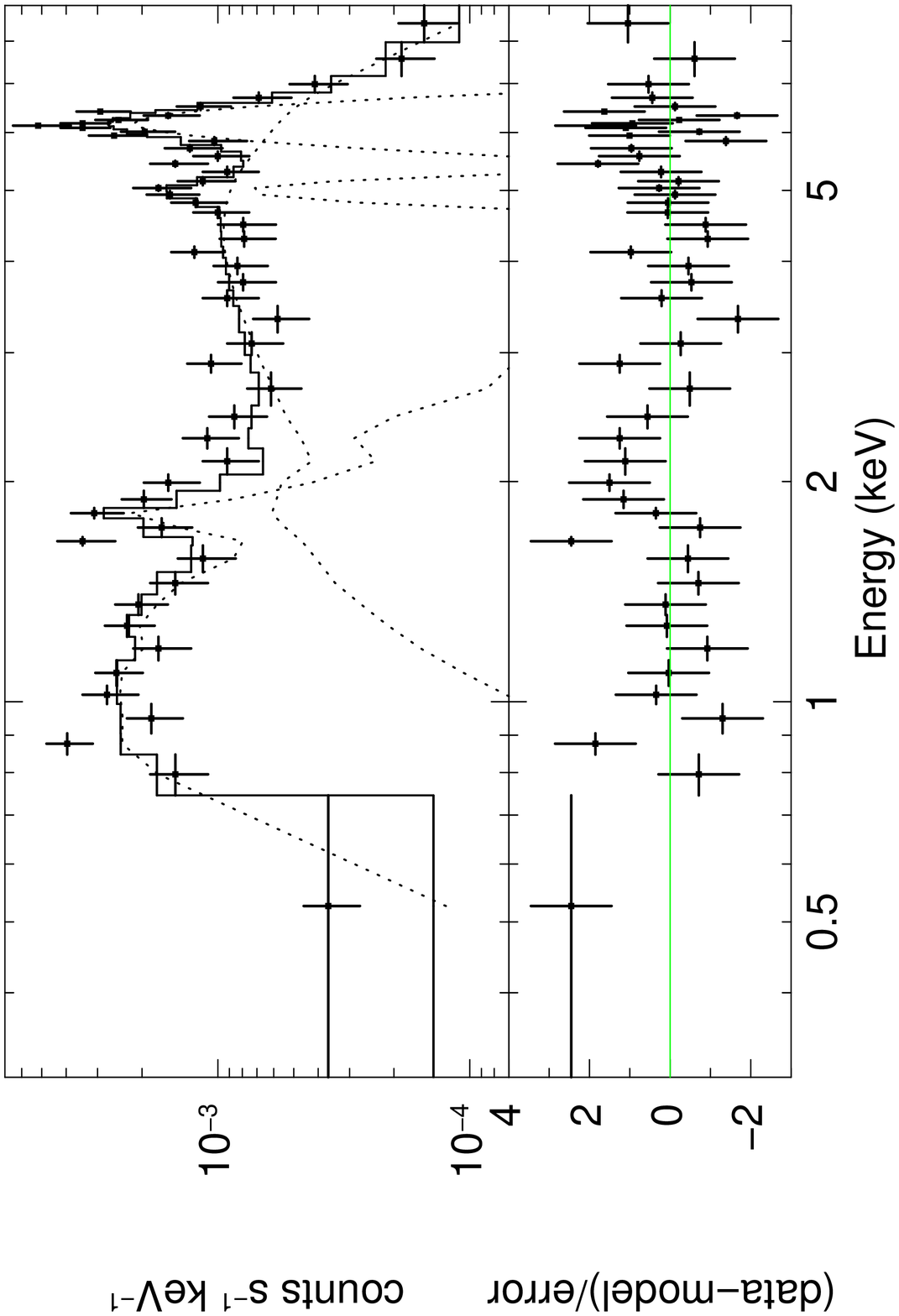}{0.2\textwidth}{\hspace{1in}(b)}
          \hspace{-1in}\rotatefig{270}{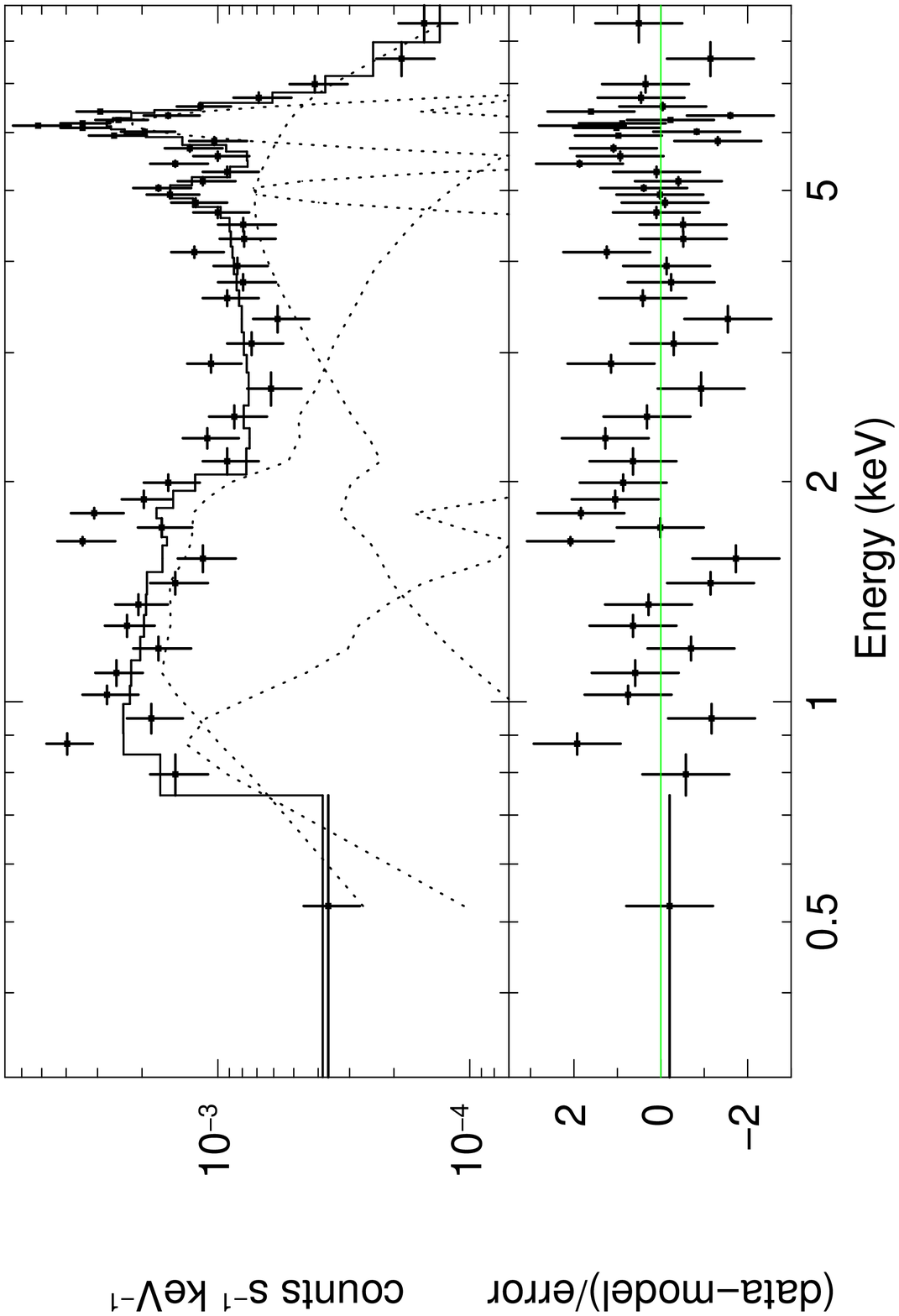}{0.2\textwidth}{\hspace{1in}(c)}
          }
\caption{Spectrum of Mrk 78 nucleus.  The model fits shown all include Galactic obscuration, a power-law component and two Gaussian lines.  In addition, (a) includes one photoionization component, (b) includes additional obscuration associated with Mrk 78, and one thermal component, and (c) includes two thermal components.  The best fit parameters are shown in Table \ref{tab:central}.
\label{fig:central}}
\end{figure*}

The spectrum of the nuclear region of Mrk 78 is shown in Figure \ref{fig:central}.  At energies $>2$ keV, the spectrum is well described by a hard power-law component and two Gaussian lines.  The strongest line has a central energy of $6.44\pm0.05$ keV, the energy of neutral Fe K$\alpha$. The hard power-law and strong iron emission is typical of obscured AGN. If this line is excluded from the model, $\chi_{\nu}^2=2.8$ and the model is ruled out with $>6\sigma$ confidence. \par
Visual inspection of the residuals suggests that there may also be a second Gaussian line at approximately 5.2 keV.  When fitting just the $3-8$ keV band, including this additional line reduces $\chi^2_{\nu}$ from 1.30 to 1.14 and increases the null hypothesis probability ($p_{\mathrm{null}}$) from 15\% to 29\%.  The central energy of this line is consistent with the 5.17 keV line seen reported in the Compton-thick AGN in NGC 7212 \citep{jones20}.  It was suggested this line could result from He-like Vanadium formed by cosmic spallation, although it is debated whether Vanadium K$\alpha$ can be strong compared to other spallation lines and it is debated to what extent spallation may occur in AGN (\citealt{skibo97}; \citealt{scott05}; \citealt{turner10}; \citealt{gallo19}).  We note that some structure remains in the residuals around $5-6$ keV, suggesting additional emission lines or non-Gaussian line profiles may be present, but additional X-ray observations would be required to test these possibilities.\par
Our power-law fits of the AGN continuum should not be taken too literally, since Mrk 78 has been shown to be close to Compton-thick based on X-ray broadband spectral fitting using \textit{NuSTAR} and \textit{XMM} observations \citep{zhao20}.  \citet{zhao20} find that the Mrk 78 nucleus exhibits a line-of-sight column density of log($N_{\mathrm{H}}$)$\approx23.8-23.9$ and that its broadband X-ray spectrum is well-fit by the combination of a thermal component (\texttt{mekal}, \citealt{mewe85}), an absorbed intrinsic cutoff power-law spectrum, a reprocessed component including scattering and fluorescent lines, and leaked unabsorbed intrinsic continuum.  They find that the thermal component dominates above the AGN/torus components below 1.5 keV.  Given that neither \textit{XMM} nor \textit{NuSTAR} can resolve the nuclear and extended emission in Mrk 78, we expect that the AGN/torus emission should be even less dominant in \textit{Chandra} spectra of the extended emission $>1^{\prime\prime}$ from the nucleus.  Therefore, our spectral fitting results of $0.3-2$ keV extended emission should not be significantly impacted by our simple treatment of the AGN continuum.  While our spectral fitting results for the nuclear region could be improved by including \textit{NuSTAR} data, the focus of this paper is the extended emission so we leave this to future work.

\begin{table*}
\begin{minipage}{\textwidth}
\centering
\footnotesize
\caption{Nuclear Region Spectral Results}
\begin{tabular}{l|l|l} \hline \hline
One Photoionization + Power-law & Obscuration * (One Thermal + Power-law) & Two Thermal + Power-law \\
\hline
log($U$)=$1.27^{+0.23}_{-0.22}$ & $N_{\mathrm{H}}$=$7.6\pm1.6\times10^{21}$ cm$^{-2}$& kT$_1$=$0.84^{+0.24}_{-0.22}$ keV \\
log($N_{\mathrm{H}}$)=$22.27^{+0.90}_{-1.94}$ & kT$_1$=$0.69^{+0.12}_{-0.13}$ keV & log($f_{0.3-2, \mathrm{apec1}}$)=$-14.33^{+0.19}_{-0.33}$ \\
log($f_{0.3-2, \mathrm{CLOUDY}}$)=$-13.75^{+0.09}_{-0.14}$ & log($f_{0.3-2, \mathrm{apec}}$)=$-12.93^{+0.16}_{-0.19}$ & kT$_2>3.09$ keV \\
$\Gamma$=$-0.64^{+0.59}_{-0.42}$ & $\Gamma$=$-0.63^{+0.44}_{-0.38}$& log($f_{0.3-2, \mathrm{apec2}}$)=$-13.84^{+0.09}_{-0.07}$ \\
Norm$_{\mathrm{PL}}$=$9.1^{+11.2}_{-4.4}\times10^{-7}$  & Norm$_{\mathrm{PL}}$=$1.01^{+0.80}_{0.44}\times10^{-6}$& $\Gamma$=$-1.47^{+0.71}_{-0.99}$\\
$E_{\mathrm{line},1}$=$6.44^{+0.06}_{-0.05}$  & $E_{\mathrm{line},1}$=$6.43^{+0.06}_{-0.05}$ & Norm$_{\mathrm{PL}}$=$1.9^{+4.6}_{-1.6}\times10^{-7}$\\
$\sigma_{\mathrm{line},1}$=$0.24^{+0.10}_{-0.07}$ & $\sigma_{\mathrm{line},1}$=$0.23^{+0.09}_{-0.07}$ & $E_{\mathrm{line},1}$=$6.43^{+0.06}_{-0.05}$ \\
Norm$_{\mathrm{line},1}$=$7.3^{+1.9}_{-1.7}\times10^{-6}$  & Norm$_{\mathrm{line},1}$=$7.2^{+1.8}_{-1.7}\times10^{-6}$ & $\sigma_{\mathrm{line},1}$=$0.21^{+0.09}_{-0.21}$\\
$E_{\mathrm{line},2}$=$5.18\pm0.11$  & $E_{\mathrm{line},2}$=$5.18\pm0.11$ & Norm$_{\mathrm{line},1}$=$6.9^{+2.0}_{-4.1}\times10^{-6}$\\
$\sigma_{\mathrm{line},2}<0.36$  & $\sigma_{\mathrm{line},2}<0.34$& $E_{\mathrm{line},2}$=$5.35^{+1.17}_{-0.27}$\\
Norm$_{\mathrm{line},2}$=$8.3^{+7.9}_{-5.5}\times10^{-7}$ & Norm$_{\mathrm{line},2}$=$7.8^{+7.5}_{-5.2}\times10^{-7}$& $\sigma_{\mathrm{line},2}<0.34$ \\
&  & Norm$_{\mathrm{line},2}$=$1.5^{+3.6}_{-1.1}\times10^{-6}$\\
$\chi^2$/DOF=57/41  & $\chi^2$/DOF=56/41  & $\chi^2$/DOF=51/40\\
$p_{\mathrm{null}}$=0.053 & $p_{\mathrm{null}}$=0.061 & $p_{\mathrm{null}}$=0.105 \\
\hline \hline
\multicolumn{3}{p{6.0in}}{\T Notes: All models include Galactic absorption with column density $N_{\mathrm{H,Gal}}=4.1\times10^{20}$ cm$^{-2}$ and two Gaussian lines.  The best-fit value for the equivalent width of the Gaussian line at $5.18$ keV was approximately $0.11$ keV in all three fits.
} \\
\end{tabular}
\label{tab:central}
\end{minipage}
\end{table*}

\subsection{Extended soft X-ray emission}
\label{sec:spectralext}
In studying the spectrum of the extended soft X-ray emission in Mrk 78, we first test whether the spectra within the Eastern and Western pie sectors along the bi-cone axis shown in Figure \ref{fig:piespectralregions} are consistent with one another.  Fitting the $0.3-2$ keV emission of the East and West regions independently with combinations of two thermal or two photoionization models, we find that the best-fit parameters for the regions are consistent within 1$\sigma$ errors, except for the overall normalization.   Thus, in the remainder of our analysis, we combine the spectra from the East and West sides of Mrk 78 to maximize the spectral statistics.  \par
While it would be preferable to treat the two sides independently throughout our spectral analysis due to the morphological differences they exhibit, since we only detect about 800 total counts in the $0.3-2$ keV band from the extended emission, we cannot simultaneously split this X-ray emission between East and West and as a function of radius.  We note that when we combine the Eastern and Western X-ray emission, both sides contribute a comparable number of counts to the inner extended emission (between $1-2^{\prime\prime}$ from the nucleus), whereas the Western side contributes 80\% of the counts to the combined outer extended emission (between $2-4.5^{\prime\prime}$ from the nucleus).  The $L$[OIII]/$L_X$ ratios of the East and West sides are are comparable, with the inner extended emission on both sides exhibiting higher ratios than the outer extended emission.  
\par

\begin{figure}
    \centering
    \includegraphics[angle=270,width=0.45\textwidth]{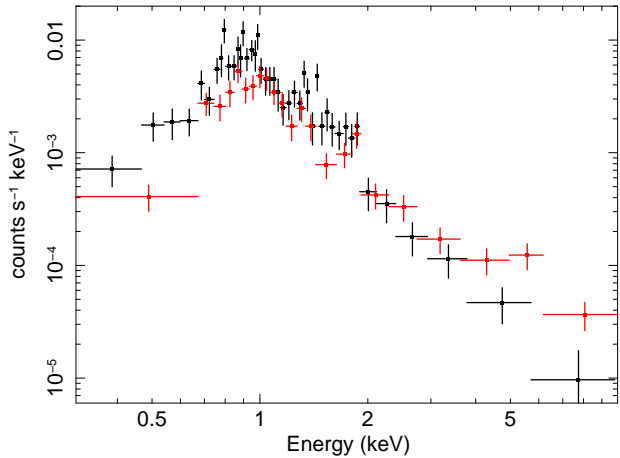}
    \caption{Spectrum of the inner extended emission is shown in red, while that of the outer extended emission is shown in black.}
    \label{fig:inneroutercomparison}
\end{figure}

Figure \ref{fig:inneroutercomparison} compares the spectra of the inner and outer pie regions shown in Figure \ref{fig:piespectralregions}.  The two spectra are well-matched in flux between $1-1.2$ keV but show significant differences.  The inner region, shown in red, exhibits lower emission in the $0.3-1$ keV band, different spectral features between $1.2-2$ keV, and higher emission between $4-8$ keV.  The latter is due to the fact that the inner region is closer to the AGN, and thus a larger fraction (about 2\%) of the counts from the hard X-ray source extend into this region due to the \textit{Chandra} PSF.   
\par

\begin{figure*}
\gridline{\hspace{-1in}\rotatefig{270}{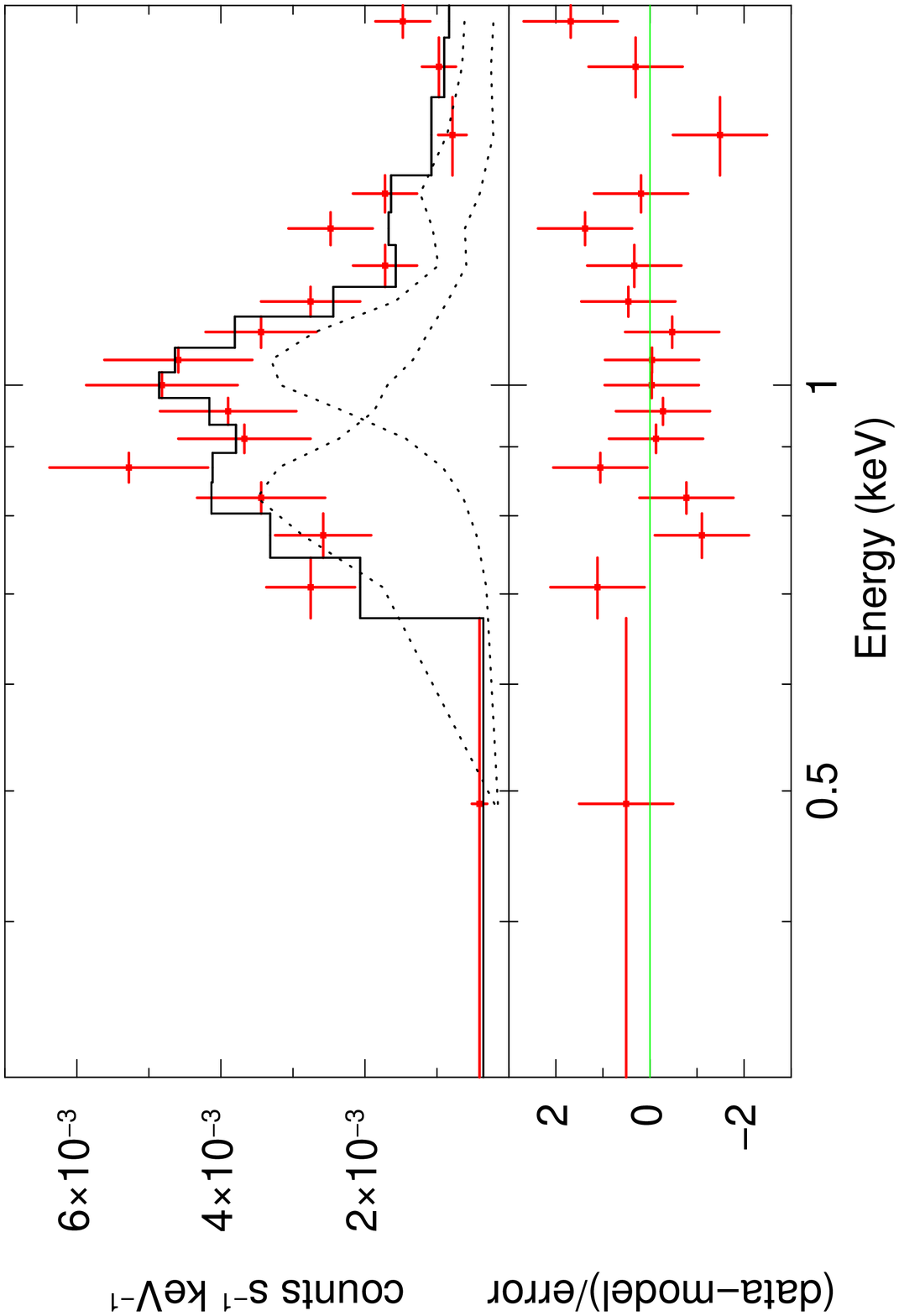}{0.3\textwidth}{\hspace{1.5in}(a)}
          \hspace{-1in}\rotatefig{270}{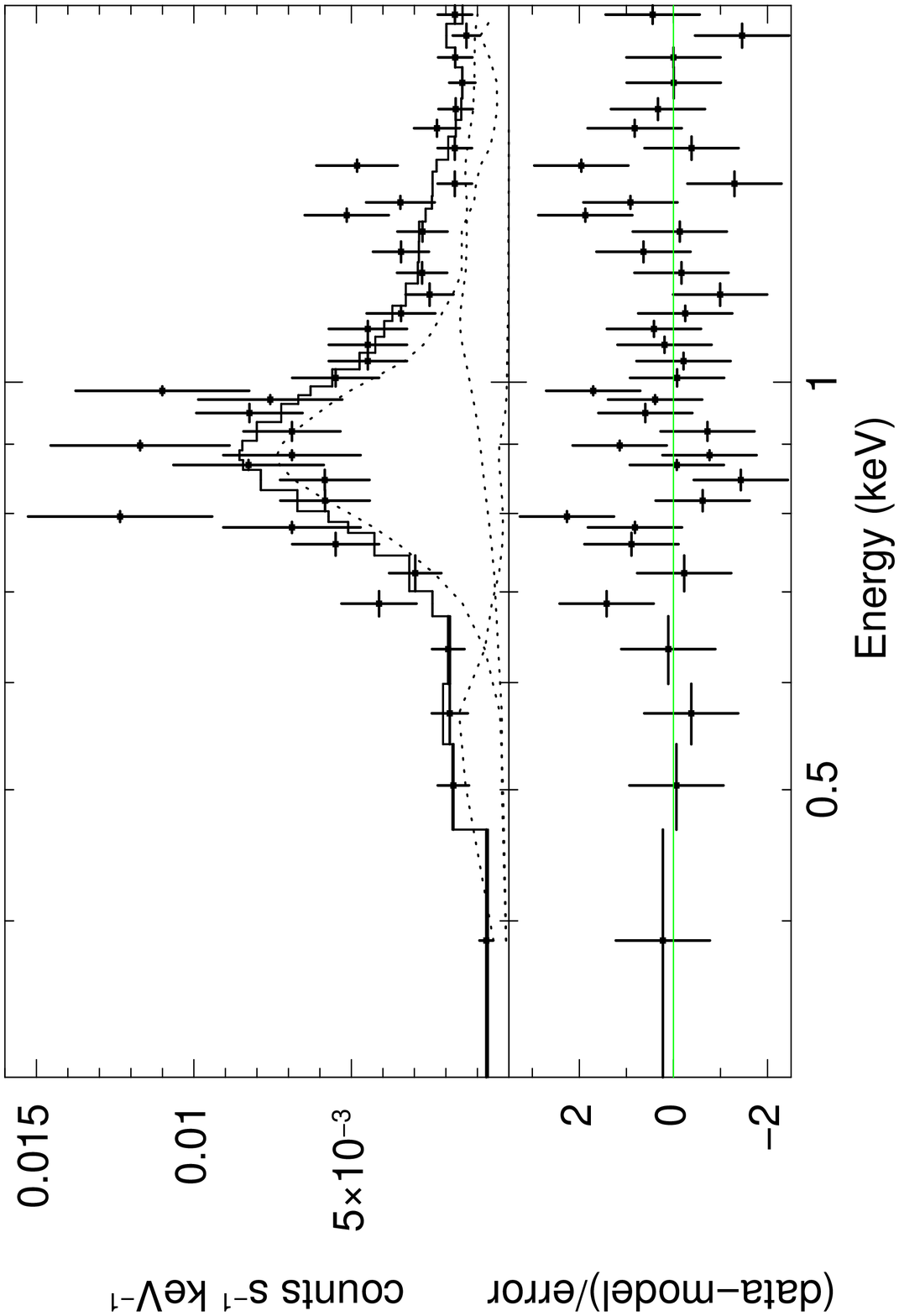}{0.3\textwidth}{\hspace{1.5in}(b)}
          }
\gridline{\hspace{-1in}\rotatefig{270}{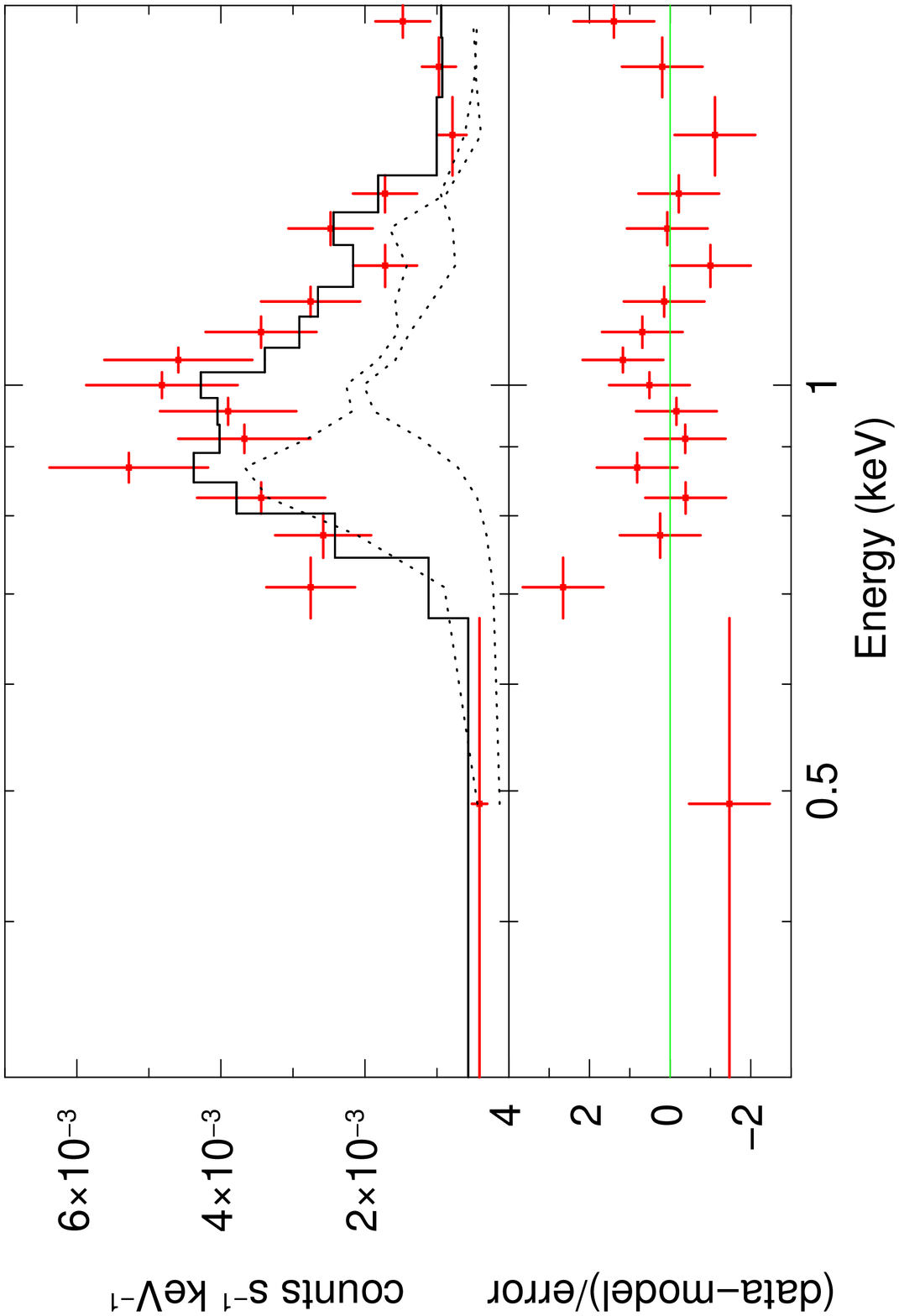}{0.3\textwidth}{\hspace{1.5in}(c) }
          \hspace{-1in}\rotatefig{270}{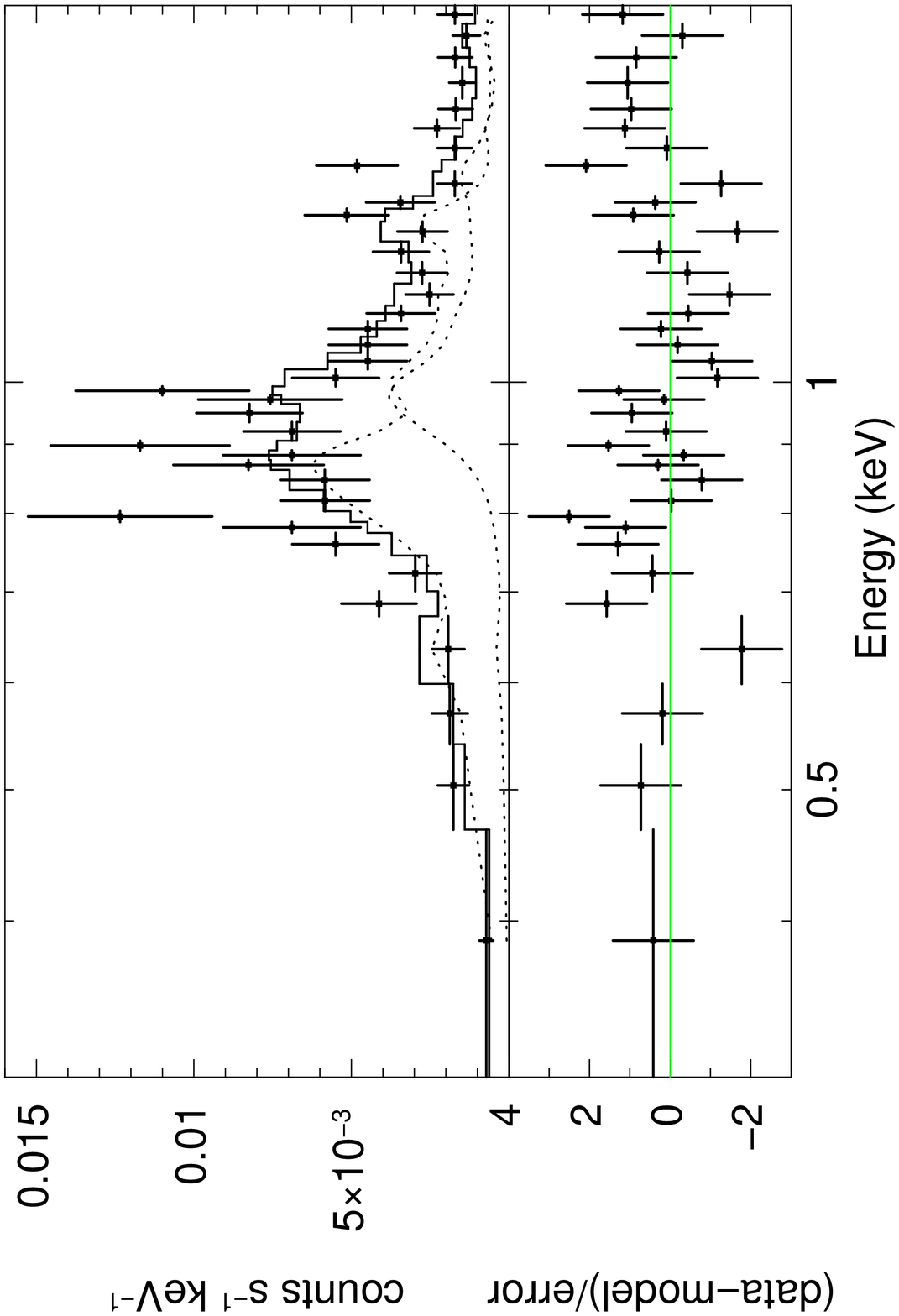}{0.3\textwidth}{\hspace{1.5in}(d)}
          }
\gridline{\hspace{-1in}\rotatefig{270}{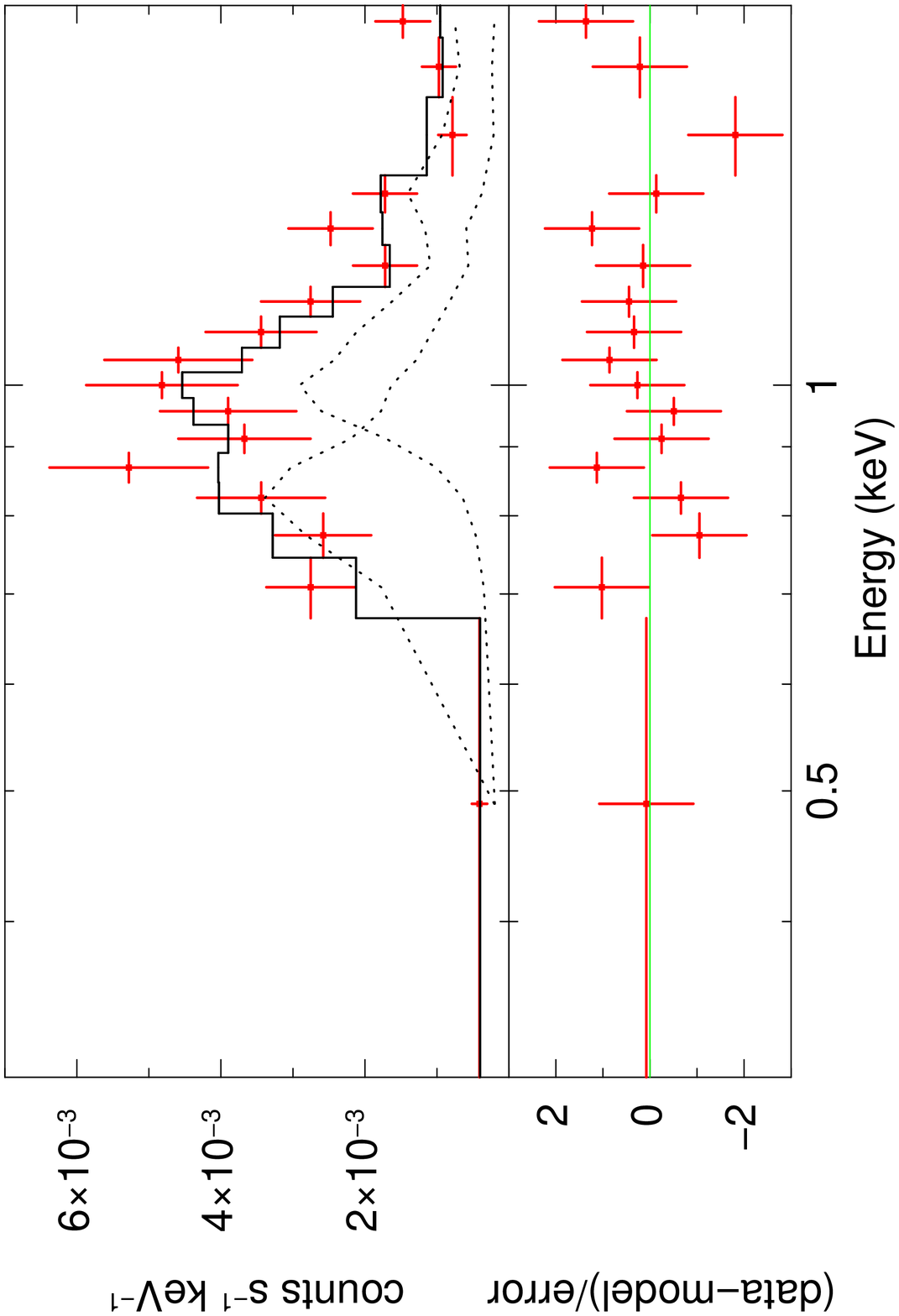}{0.3\textwidth}{\hspace{1.5in}(e) }
          \hspace{-1in}\rotatefig{270}{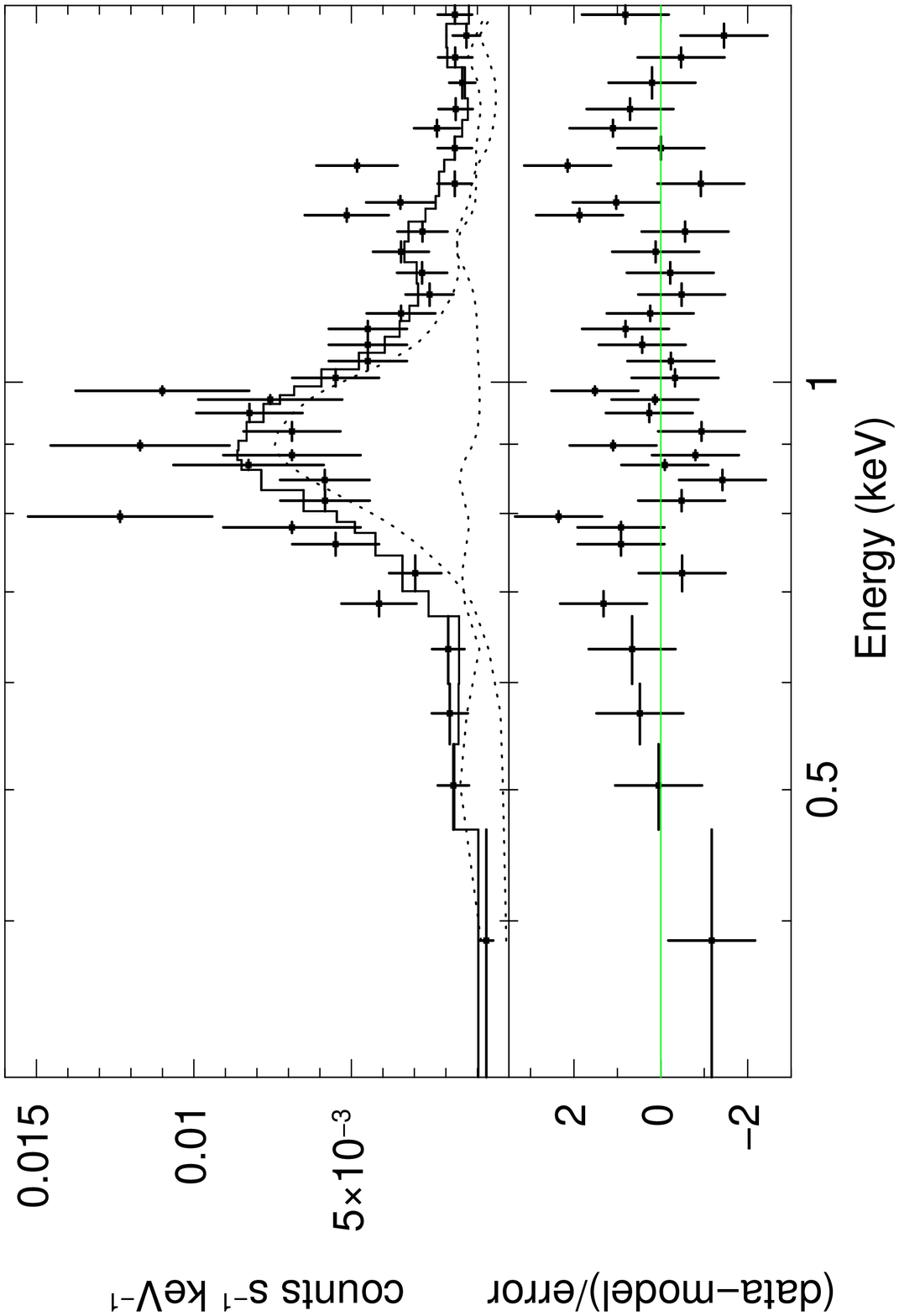}{0.3\textwidth}{\hspace{1.5in}(f)}
          }
\caption{Spectra of inner and outer extended emission regions are shown in red and black, respectively.  The model fits shown include two thermal components in (a), three thermal components in (b), two photoionization components in (c) and (d), and a mix of one photoionization and one thermal component in (e) and (f).  Best fit parameters are provided in Table \ref{tab:spectra}.
\label{fig:innerouterspectra}}
\end{figure*}

Given the visible differences between the inner and outer region spectra, we fit them independently.  We present the results based on fitting the $0.3-2$ keV band in Table \ref{tab:spectra}.  At these energies, the contribution from the hard X-ray source is minimal.  We find consistent results if we fit the $0.3-8$ keV band, fixing the normalizations of the power-law component and the 6.4 and 5.2 keV lines to the expected contributions from the nuclear hard X-ray source based on the \textit{Chandra} PSF using the CIAO \texttt{psfFrac} tool. 

\par
Neither the inner nor outer $0.3-2$ keV spectra can be well-fit using only a single thermal or photoionization model subject only to Galactic obscuration.  The inner region spectrum can be well described with either a single thermal or photoionization model subject to additional, presumably local to Mrk 78, obscuration of $N_{\mathrm{H}}\approx4.5\times10^{21}$ cm$^{-2}$, or any two component combination of photoionization and/or thermal models without additional obscuration.  

\par
The outer region spectrum can also be described well by a combination of either two photoionization models or one photoionization plus one thermal model.  In the case of the two photoionization models, the ionization parameters of the outer region are slightly lower than for the inner region, and the photoionization component with relatively lower ionization parameter is more dominant than for the inner region. Using the $2-10$ keV intrinsic luminosity from \citet{zhao20} and the AGN SED from \citet{elvis94}, we estimate $Q_{\mathrm{ion}}\approx2\times10^{55}$ ionizing photons per second; thus, the lower ionization parameter log$U\approx1$ corresponds to $n_e\sim1$ cm$^{-3}$, while the higher ionization parameter log$U\approx2$ corresponds to $n_e\sim0.1$ cm$^{-3}$.  Note that for this X-ray gas to be in pressure balance with the optical line-emitting gas with $n_e\sim10^{2}-10^{3}$ cm$^{-3}$ and $T\sim10^4$ K \citepalias{whittle05}, a temperature of $\sim10^{7}$ K would be required, which is reasonable for hot X-ray gas.  \par
When adopting the mixture of one photoionization plus one thermal component, the ionization parameter of the outer region is much lower and its temperature slightly higher than for the inner region; while for the inner region, the photoionization and thermal component contribute the same amount of flux, in the outer region, the photoionization component dominates. The low ionization parameter preferred by this model fit would indicate a higher density of $n_e\sim300$ cm$^{-3}$.  However, note that it is possible to produce an acceptable fit to the outer region with $\chi^2=46$ for 33 degrees of freedom (corresponding to $p_{\mathrm{null}}=0.070$) with an ionization parameter (log$U\approx1.86$) and thermal temperature($kt\approx0.75$ keV) that are consistent with the values for the inner region; in this case, the thermal component dominates the flux in the outer region. A model with a single photoionized component subject to obscuration is excluded with $>95$\% probability for the outer region.

\par 
In order to produce acceptable fits to the outer region using only thermal models, either three thermal without additional obscuration, or two thermal models obscured by $N_{\mathrm{H}}\approx3.5\times10^{21}$ cm$^{-2}$ in excess of the Galactic values are required.  For both these sets of models for the outer region, the lowest temperature thermal component is the most dominant and it has a significantly lower temperature than any of the best-fit thermal components for the inner region.
Given that more components are required to fit the spectrum with a thermal-only model compared to photoionization-only or mixed models, and that we know that an AGN photoionizing source is present, it seems unlikely that all the emission in the outer region originates in thermal shocks.  

\par
Figure \ref{fig:innerouterspectra} displays the model fits for the inner and outer region based on pure combinations of thermal or photoionization models.  The observed and intrinsic fluxes and luminosities of the $0.3-2$ keV emission derived using all the aforementioned models are provided in Table \ref{tab:fluxlum}.
\par

\begin{table*}
\begin{minipage}{\textwidth}
\centering
\footnotesize
\caption{Extended Region Spectral Results}
\begin{tabular}{ll|ll} \hline \hline
\multicolumn{2}{l|}{Inner: Two Photoionization}  & \multicolumn{2}{l}{Outer: Two Photoionization} \\
\hline
log($U_1$)=$1.00^{+0.21}_{-0.15}$ &  log($U_2$)=$1.93^{+>0.07}_{-0.15}$ & log($U_1$)=$0.75^{+0.10}_{-0.18}$ &  log($U_2$)=$1.87^{+>0.13}_{-0.28}$ \\
log($N_{\mathrm{H,1}}$)=$19.87^{+0.64}_{->0.87}$ & log($N_{\mathrm{H,2}}$)$<21.56$ & log($N_{\mathrm{H,1}}$)=$19.79^{+0.56}_{-0.52}$ & log($N_{\mathrm{H,2}}$)=$19-23.5$\\
log($f_{0.3-2,1}$)=$-13.65^{+0.11}_{-0.14}$ & log($f_{0.3-2,2}$)=$-14.06^{+0.16}_{-0.40}$ & log($f_{0.3-2,1}$)=$-13.28^{+0.07}_{-0.05}$ & log($f_{0.3-2,2}$)=$-13.88^{+0.14}_{-0.17}$\\
$\chi^2$/DOF=17/12 & $p_{\mathrm{null}}$=0.141 & $\chi^2$/DOF=42/32 & $p_{\mathrm{null}}$=0.116 \\
\hline
\multicolumn{2}{l|}{Inner: Two Thermal} & \multicolumn{2}{l}{Outer: Three Thermal} \\
\hline
kT$_1$=$0.67^{+0.16}_{-0.21}$ keV & kT$_2$=$1.60^{+0.79}_{-0.27}$ keV & kT$_1$=$0.12^{+0.08}_{-0.23}$ keV & kT$_2$=$0.84\pm0.09$ keV \\
log($f_{0.3-2,1}$)=$-13.85^{+0.10}_{-0.13}$ & log($f_{0.3-2,2}$)=$-13.93^{+0.10}_{-0.14}$ & log($f_{0.3-2,1}$)=$-13.45^{+0.18}_{-0.23}$ & log($f_{0.3-2,2}$)=$-13.59^{+0.07}_{-0.09}$ \\
$\chi^2$/DOF=12/13 & $p_{\mathrm{null}}$=0.515 & kT$_3>2.67$ keV & log($f_{0.3-2,3}$)=$-13.91^{+0.11}_{-0.17}$\\
& & $\chi^2$/DOF=32/32 & $p_{\mathrm{null}}$=0.464 \\
\hline
\multicolumn{2}{l|}{Inner: One Photoionization + One Thermal} & \multicolumn{2}{l}{Outer: One Photoionization + One Thermal} \\
\hline
log($U_1$)=$1.94^{+>0.06}_{-0.16}$ & kT$_2$=$0.65^{+0.17}_{-0.16}$ keV & log($U_1$)=$-1.41^{+0.50}_{-0.52}$ & kT$_2$=$0.90^{+0.09}_{-0.08}$ keV\\
log($N_{\mathrm{H,1}}$)$<21.34$ & log($f_{0.3-2,2}$)=$-13.86^{+0.10}_{-0.11}$ & log($N_{\mathrm{H,1}}$)=$20.86^{+1.02}_{-1.11}$ & log($f_{0.3-2,2}$)=$-13.60^{+0.07}_{-0.08}$\\
log($f_{0.3-2,1}$)=$-13.89^{+0.08}_{-0.15}$ & & log($f_{0.3-2,1}$)=$-13.28^{+0.10}_{-0.15}$ &\\
$\chi^2$/DOF=12/12 & $p_{\mathrm{null}}$= 0.419 & $\chi^2$/DOF=35/33 & $p_{\mathrm{null}}$= 0.370\\
\hline
\multicolumn{2}{l|}{Inner: Obscuration*(One Thermal)} & \multicolumn{2}{l}{Outer: Obscuration*(Two Thermal)} \\
\hline
kT=$0.58^{+0.14}_{-0.31}$ keV & log($f_{0.3-2}$)=$-13.00^{+0.72}_{-0.17}$ & kT$_1$=$0.10\pm0.04$ keV & kT$_2$=$0.82^{+0.10}_{-0.39}$ keV \\
$N_{\mathrm{H},z}$=$4.5^{+2.7}_{-1.0}\times10^{21}$ cm$^{-2}$ & & log($f_{0.3-2,1}$)=$-11.99^{+0.46}_{-0.48}$ & log($f_{0.3-2,2}$)=$-13.09^{+0.23}_{-0.12}$ \\
$\chi^2$/DOF=20/14 & $p_{\mathrm{null}}$=0.143 & $N_{\mathrm{H},z}$=$3.5^{+1.6}_{-1.2}\times10^{21}$ cm$^{-2}$  & \\
& & $\chi^2$/DOF=44/33 & $p_{\mathrm{null}}$=0.092 \\
\hline
\multicolumn{2}{l|}{Inner: Obscuration*(One Photoionization)} & \\
\hline
log($U$)=$0.20^{+0.14}_{-0.16}$ & log($f_{0.3-2}$)=$-12.48^{+0.22}_{-0.21}$ & & \\
log($N_{\mathrm{H}}$)=$19.66^{+0.46}_{->0.66}$ & $N_{\mathrm{H},z}$=$4.5^{+1.2}_{-1.0}\times10^{21}$ cm$^{-2}$\\
$\chi^2$/DOF=18/13 & $p_{\mathrm{null}}$=0.147 & & \\
\hline \hline
\multicolumn{4}{p{6.0in}}{\T Notes: All models include Galactic absorption with column density $N_{\mathrm{H,Gal}}=4.1\times10^{20}$ cm$^{-2}$.
} \\
\end{tabular}
\label{tab:spectra}
\end{minipage}
\end{table*}





We also fit the total $0.3-2$ keV extended emission in an annular region between $1-4.5^{\prime\prime}$ from the nucleus.  A combination of at least three thermal or photoionization models are required to fit this spectrum, even if obscuration in excess of Galactic obscuration is included.  The only three-component combination that is excluded with $>95$\% confidence consists of three photoionization models, which has $p_{\mathrm{null}}=1.5$\%.  All other possible three-component combinations have a null hypothesis probability $>38$\%.  However, it is unlikely that all the soft X-ray emission results from collisional thermal processes, since some photoionization due to the central AGN is expected.  Thus, it seems most likely that the soft X-ray emission results from a mixture of photoionization and collisional ionization, the latter of which may arise in shocks.

Overall, our spectral analysis reveals that the extended soft X-ray emission in Mrk 78 arises from a complex medium, exhibiting a range of densities and temperatures, as seen in a number of AGN extended X-ray regions (e.g. \citealt{travascio21}; \citealt{jones20}; \citealt{fischer19}; \citealt{maksym19}; \citealt{paggi12}).  The photoionization models indicate that the ionization parameter of the photoionized emission is either roughly constant or decreases with distance from the AGN.  Since $U\propto n_e^{-1} r^{-2}$, this trend implies that the electron density drops off as $r^{-2}$ or less steeply.  The optical line emitting gas, which is found to be primarily photoionized, similarly exhibits a decreasing ionization parameter with increasing radius \citepalias{whittle05}.  The thermal models indicate that the low temperature components become more dominant as the radius increases, and the temperature of the collisionally-ionized emission either decreases or remains constant with distance from the AGN. 


\subsubsection{900 eV emission}

\begin{figure}
\centering
\includegraphics[angle=270,width=0.45\textwidth]{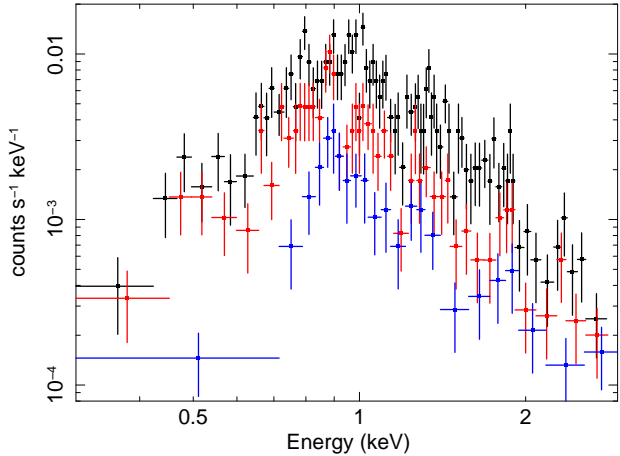}
\caption{$0.3-3$ keV spectra of the inner and outer sets of 900 eV emission regions (displayed in Figure \ref{fig:neiximages}) are shown in blue and red, respectively.  The spectrum of the remaining extended emission (excluding the 900 eV emission regions) $>1^{\prime\prime}$ from the nucleus is shown in black.}
\label{fig:neixspectra}
\end{figure}

In Figure \ref{fig:neixspectra}, we compare the spectra extracted from the regions with 900 eV emission shown in Figure \ref{fig:neiximages} to the remainder of the extended emission located $>1^{\prime\prime}$ from the nucleus.  These spectra are binned with a minimum of only 5 counts per bin, in order to maintain fine energy resolution so that differences between the spectra are more obvious.  In order to determine the element species that can account for the differences between the spectra of the 900 eV regions and the remainder of the extended emission, we attempt to fit the $0.3-2$ keV spectra with a combination of Gaussian lines at fixed energies corresponding to species typically seen in ionized AGN bicones (\citealt{paggi12}; \citealt{maksym19}; \citealt{jones20}).  The complete list of line energies used can be seen in Table \ref{tab:900spectra}.  \par
We perform one set of fits fixing all the line widths to $\sigma=50$ eV.  We perform a second set of fits leaving the line width as a free parameter but tieing all the line widths to each other since otherwise there are too many free parameters for meaningful fits.  Given the low number of counts per bin, we use the L-statistic to perform the fitting, but also report the chi-square statistic in Table \ref{tab:900spectra} to provide a rough measure of the goodness of fit.  We calculate the 90\% confidence errors on the free parameters using Monte Carlo Markov chains with 10,000 steps.  If a particular line is found to have a normalization upper limit that is $<10^{-8}$, we fix that line normalization to zero and perform the fit again. \par
\begin{figure*}
\gridline{\hspace{-1in}\rotatefig{270}{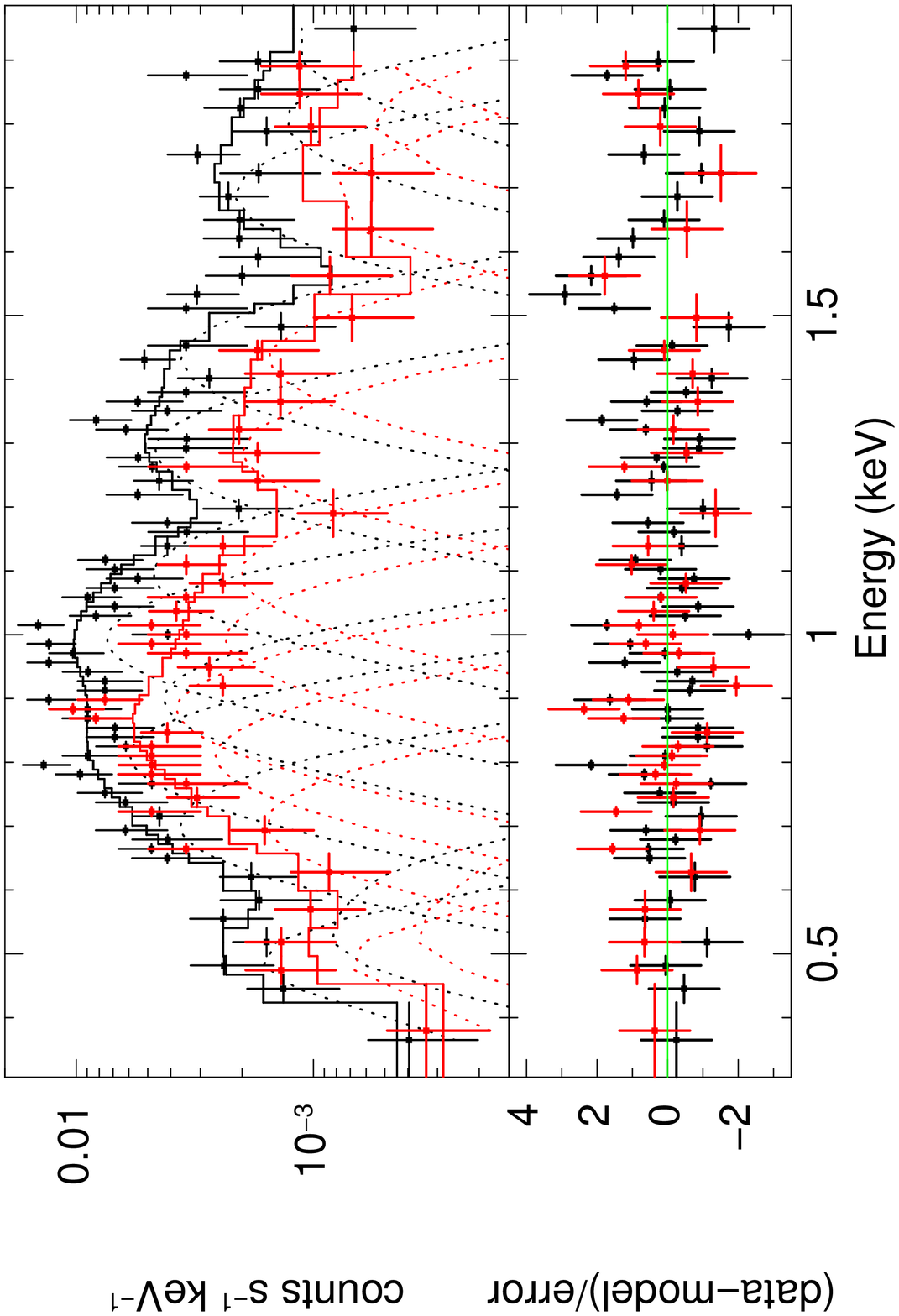}{0.3\textwidth}{\hspace{1.5in}(a)}
          \hspace{-1in}\rotatefig{270}{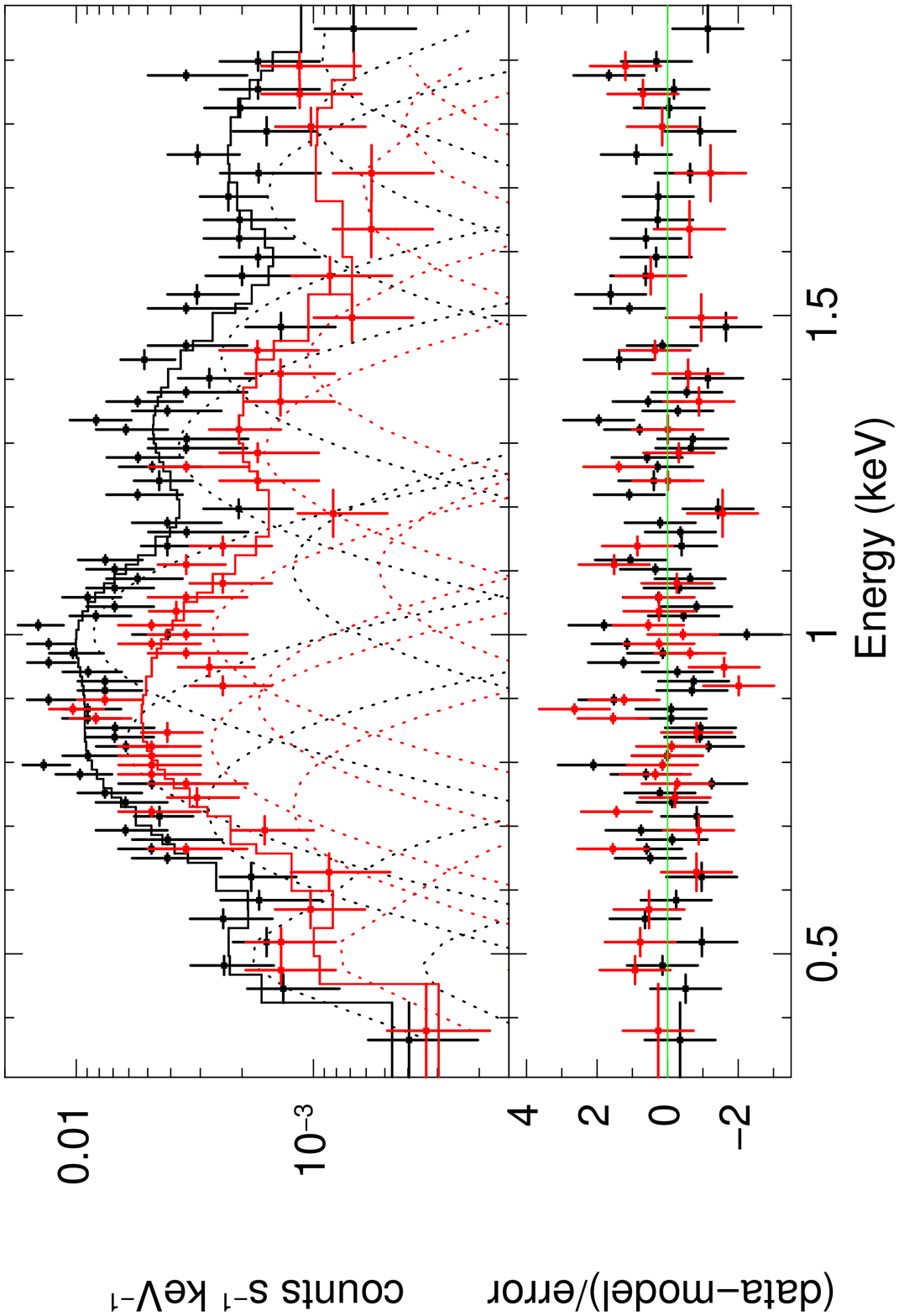}{0.3\textwidth}{\hspace{1.5in}(b)}
          }
\caption{The spectrum of the outer 900 eV emission regions shown in red in Figure \ref{fig:neiximages} is shown in red, while the spectrum of the remainder of the extended emission at $>1^{\prime\prime}$ from the nucleus is shown in black.  The two spectra are fit jointly with a combination of Gaussian lines subject to absorption by a Galactic column density of $4.1\times10^{20}$ cm$^{-2}$.  A constant scale factor adjusts the overall normalization of the two spectra.  In (a) all lines have a fixed width of $\sigma=50$ eV, and in (b) the line widths are tied together but the value is allowed to vary. Best fit parameters are provided in Table \ref{tab:900spectra}.
\label{fig:jointspectra}}
\end{figure*}

We first fit the spectra of the outer 900 eV emission regions and the remainder of the extended emission jointly, allowing for a scaling factor that adjusts the overall normalization but keeps the relative normalizations between different Gaussian lines the same between the two spectra.  As can be seen in Figure \ref{fig:jointspectra}, the two spectra exhibit significantly different residuals to this joint best fit between approximately 0.85 and 1.0 keV. Therefore, we performed a second fit, allowing the normalizations of Gaussian lines in the two spectra in this energy range to be independent of each other. This results in an improved fit, with a lower L-statistic and flatter residuals.  Table \ref{tab:900spectra} provides results for the outer 900 eV emission regions, shown in red in Figure \ref{fig:neiximages}, and the remainder of the extended emission located $>1^{\prime\prime}$ from the nucleus.  Regardless of whether the line width is fixed to 50 eV or left free to vary, the spectrum of the outer 900 eV regions are consistent with enhanced emission around 0.905 keV (the line energy of Ne IX) and lower emission around 1.022 keV (the line energy of Ne X).  \par
The spectra of the outer 900 eV emission regions and the remainder of the extended emission have $300\pm17$ and $664\pm26$ net counts in the $0.3-2$ keV band, respectively\footnote{Note that the sum of the $0.3-2$ keV counts in the 900 eV regions and the remainder of the extended emission is slightly higher than the sum of the counts in the inner and outer extended emission regions shown in Figure \ref{fig:piespectralregions} because parts of the 900 eV emission regions lie outside the boundaries of the inner/outer extended emission regions.}.  Although we analyzed the spectra of the Eastern and Western outer 900 eV regions jointly, we note that there are some differences between the morphology of the 900 eV emission as compared to the [OIII] and radio emission on the Eastern and Western sides.  As discussed in \S\ref{sec:morphcomparison}, the 900 eV emission associated with the Eastern knot, XE900-1, lies farther away from the nucleus compared the to the peak of the [OIII] and radio emission on the Eastern side.  Instead, in the Western arc, the 900 eV peaks (XW900-1 and XW900-2) are spatially coincident with bright [OIII] and radio emission rather than lying on the outskirts of it.  More X-ray data would be required to analyze the spectra of the 900 eV emission in these Eastern and Western regions independently. 
\par
The spectrum of the inner 900 eV regions near the nucleus shown in cyan in Figure \ref{fig:neiximages} only have $117\pm11$ combined net counts in the $0.3-2$ keV band.  Modeling the inner 900 eV emission spectrum is further complicated by the fact that the power-law component provides a non-negligible contribution to the $0.3-2$ keV band, requiring additional free parameters in the model.  Therefore, we were not able to perform detailed line modeling for the inner 900 eV emission regions. \par
Nonetheless, we can visually compare their combined spectrum (in blue) in Figure \ref{fig:neixspectra} with the spectra of the outer 900 eV emission regions (in red) and the remainder of the extended emission (in black).  The inner 900 eV emission regions appear to exhibit the same excess at the Ne IX energy as the outer 900 eV emission regions.  However, more X-ray data would be required to make a robust determination.  

\begin{table*}
\begin{minipage}{\textwidth}
\centering
\footnotesize
\caption{900 eV Emission Spectral Results}
\begin{tabular}{lccccccc} \hline \hline
& & Joint & 900 eV Regions & Remainder & Joint & 900 eV Regions & Remainder \\
\hline
& & \multicolumn{3}{c}{Fixed $\sigma$} & \multicolumn{3}{c}{Variable $\sigma$} \\
\hline
Species & Rest  & Normalization & Normalization & Normalization & Normalization & Normalization & Normalization \\
& Energy (keV) & (1e-6) & (1e-6) & (1e-6) & (1e-6) & (1e-6) & (1e-6) \\
\hline
C \scriptsize{V} He$\gamma$ & 0.371 & 12.15$^{+12.97}_{-5.72}$ & \multicolumn{2}{c}{12.65$^{+2.82}_{-2.17}$} & ... & \multicolumn{2}{c}{...} \\
N \scriptsize{VI} triplet & 0.426 & ... & \multicolumn{2}{c}{...} & ... & \multicolumn{2}{c}{...} \\
C \scriptsize{IV} Ly$\beta$ & 0.436 & ... & \multicolumn{2}{c}{...} & 11.24$^{+30.35}_{-10.27}$ & \multicolumn{2}{c}{7.61$^{+49.24}_{-5.68}$} \\
N \scriptsize{VII} Ly$\alpha$ & 0.500 & 16.03$^{+18.85}_{-4.28}$ & \multicolumn{2}{c}{16.11$^{+2.99}_{-4.83}$} & 23.60$^{+9.66}_{-20.98}$ & \multicolumn{2}{c}{25.76$^{+9.51}_{-24.16}$} \\
O \scriptsize{VII} triplet & 0.569 &  7.97$^{+1.25}_{-2.83}$ & \multicolumn{2}{c}{8.30$^{+3.02}_{-0.02}$} & ... & \multicolumn{2}{c}{...} \\
O \scriptsize{VIII} Ly$\alpha$ & 0.654 & ... & \multicolumn{2}{c}{...} &  ... & \multicolumn{2}{c}{...} \\
Fe \scriptsize{XVII} & 0.720 & 11.34$^{+0.88}_{-0.39}$ & \multicolumn{2}{c}{11.36$^{+0.05}_{-0.54}$} & 10.43$^{+4.26}_{-6.31}$ & \multicolumn{2}{c}{11.43$^{+6.31}_{-4.71}$} \\
Fe \scriptsize{XVII} & 0.826 & 8.58$^{+0.77}_{-1.75}$ & \multicolumn{2}{c}{9.28$^{+0.73}_{-0.72}$} & 14.70$^{+4.98}_{-2.74}$ & \multicolumn{2}{c}{12.87$^{+4.65}_{4.53}$} \\
Fe \scriptsize{XVIII} & 0.873 & ... & \multicolumn{2}{c}{...} & ... & \multicolumn{2}{c}{...} \\
Ni \scriptsize{XIX} & 0.884 & ... & ... & ... & ... & ... & ... \\
Ne \scriptsize{IX} & 0.905 & 5.02$^{+0.19}_{-0.42}$ & 9.13$^{+3.29}_{-0.46}$ & ... & ... & 5.62$^{+7.09}_{-5.14}$ & ...\\
Fe \scriptsize{XIX} & 0.917 & 0.87$^{+1.93}_{-0.85}$ & ... & $<1.65$ & ... & ... & ...\\
Fe \scriptsize{XIX} & 0.922 & ... & ... & 4.19$^{+0.01}_{-1.39}$ & ... & ... & ... \\
Ne \scriptsize{X} & 1.022 & 4.85$^{+0.82}_{-1.85}$ & 3.78$^{+1.31}_{-0.06}$ & 5.21$^{+0.68}_{-0.54}$ & 8.29$^{+1.07}_{-2.07}$ & 5.93$^{+3.12}_{-2.88}$ & 8.23$^{+2.33}_{-1.01}$ \\
Fe \scriptsize{XXIV} & 1.129 & 2.48$^{+0.38}_{-0.17}$ & \multicolumn{2}{c}{2.56$^{+0.52}_{-0.97}$} & 0.37$^{+1.08}_{-0.33}$ & \multicolumn{2}{c}{0.92$^{+0.40}_{-0.79}$} \\
Fe \scriptsize{XXIV} & 1.168 & ... & \multicolumn{2}{c}{...} & ... & \multicolumn{2}{c}{...} \\
Mg \scriptsize{XI} & 1.331 & 1.82$^{+0.15}_{-0.33}$ & \multicolumn{2}{c}{1.85$^{+0.43}_{-0.50}$} & 2.14$^{+0.30}_{-0.67}$ & \multicolumn{2}{c}{2.06$^{+0.66}_{-0.29}$} \\
Mg \scriptsize{XI} & 1.352 & ... & \multicolumn{2}{c}{...} & ... & \multicolumn{2}{c}{...} \\
Mg \scriptsize{XII} & 1.478 & 1.25$^{+0.37}_{-0.16}$ & \multicolumn{2}{c}{1.27$^{+0.05}_{-0.01}$} & 1.13$^{+0.49}_{-0.33}$ & \multicolumn{2}{c}{1.19$^{+0.40}_{-0.39}$} \\
Mg \scriptsize{XII} & 1.745 & 0.74$^{+0.07}_{-0.03}$ & \multicolumn{2}{c}{0.75$^{+0.01}_{-0.01}$} & 0.73$^{+0.25}_{-0.37}$ & \multicolumn{2}{c}{0.75$^{+0.26}_{-0.31}$} \\
Si \scriptsize{XIII} & 1.839 & ... & \multicolumn{2}{c}{...} & ... & \multicolumn{2}{c}{...} \\
Si \scriptsize{XIII} & 1.865 & 0.42$^{+0.01}_{-0.02}$ & \multicolumn{2}{c}{0.43$\pm0.02$} & 0.44$^{+0.31}_{-0.19}$ & \multicolumn{2}{c}{0.43$^{+0.34}_{-0.30}$} \\
Si \scriptsize{XIV} & 2.005 & 0.45$^{+0.04}_{-0.09}$ & \multicolumn{2}{c}{0.45$^{+0.01}_{-0.15}$} & 0.47$^{+0.26}_{-0.23}$ & \multicolumn{2}{c}{0.47$^{+0.17}_{-0.26}$} \\
\hline
Scale factor & & 0.460$^{+0.003}_{-0.006}$ & 0.435$^{+0.046}_{-0.047}$ & 1.0 & 0.459$^{+0.05}_{-0.042}$ & 0.437$^{+0.052}_{-0.070}$ & 1.0\\
$\sigma$ (eV) & & 50 & \multicolumn{2}{c}{50} & 87$^{+9}_{-7}$ & \multicolumn{2}{c}{82$^{+12}_{-4}$} \\
L-statistic & & 111.18 & \multicolumn{2}{c}{104.57} & 100.49 & \multicolumn{2}{c}{96.96} \\
$\chi^2$/dof & & 112.82/102 & \multicolumn{2}{c}{110.05/100} & 106.69/104 & \multicolumn{2}{c}{105.89/102} \\ 
\hline\hline
\end{tabular}
\label{tab:900spectra}
\end{minipage}
\end{table*}

\begin{table*}
\begin{minipage}{\textwidth}
\centering
\footnotesize
\caption{Spectral Region Counts, Fluxes, and Luminosities}
\begin{tabular}{llcccccc} \hline \hline
\multirow{2}{*}{Region} & \multirow{2}{*}{Model} & Net counts  & Net counts & Observed $f_{\mathrm{X}}$ & Intrinsic $f_{\mathrm{X}}$ & Observed $L_{\mathrm{X}}$ & Intrinsic $L_{\mathrm{X}}$\\
& & (0.3-2 keV) & (0.3-8 keV) & ($10^{-14}$ erg cm$^{-2}$ s$^{-1}$) & ($10^{-14}$ erg cm$^{-2}$ s$^{-1}$) & ($10^{40}$ erg s$^{-1}$) & ($10^{40}$ erg s$^{-1}$) \\
\hline
\multirow{3}{*}{Nucleus} & 1P & \multirow{3}{*}{$261\pm16$} & \multirow{3}{*}{$858\pm29$} & $1.77^{+0.15}_{-0.14}$ & $1.78^{+0.41}_{-0.49}$ & $2.48^{+0.21}_{-0.19}$ & $2.49^{+0.58}_{-0.68}$ \\
 & O*1T & & & $1.42^{+0.10}_{-0.09}$ & $11.7^{+5.2}_{-4.2}$ & $1.98^{+0.14}_{-0.12}$ & $16.4^{+7.3}_{-5.8}$\\
 & 2T & & & $1.71^{+0.12}_{-0.68}$ & $1.91^{+0.40}_{-0.33}$ & $2.39^{+0.17}_{-0.95}$ & $2.68^{+0.56}_{-0.46}$ \\ \hline
\multirow{5}{*}{Inner} & 2P & \multirow{5}{*}{$294\pm17$} & \multirow{5}{*}{$383\pm20$} & $2.43^{+0.29}_{-0.12}$ & $3.11^{+0.75}_{-0.81}$ & $3.40^{+0.41}_{-0.17}$ & $4.35^{+1.05}_{-1.13}$\\
 & 2T & & & $2.19^{+0.18}_{-0.11}$ & $2.59\pm0.48$ & $3.07^{+0.25}_{-0.15}$ & $3.62\pm0.67$\\
 & 1P+1T & & & $2.25^{+0.16}_{-0.14}$ & $2.67^{+0.46}_{-0.47}$ & $3.15^{+0.22}_{-0.19}$ & $3.74^{+0.65}_{-0.66}$\\
 & O*1T & & & $1.91^{+0.11}_{-0.13}$ & $10.0^{+43.7}_{-3.2}$ & $2.68^{+0.15}_{-0.19}$ & $14.0^{+61.2}_{-4.5}$\\
 & O*1P & & & $2.15^{+0.25}_{-0.18}$ & $33.1^{+21.8}_{-12.7}$ & $3.01^{+0.35}_{-0.25}$ & $46.3^{+30.6}_{-17.8}$\\ \hline
\multirow{4}{*}{Outer} & 2P & \multirow{4}{*}{$498\pm22$} & \multirow{4}{*}{$549\pm24$} & $4.82^{+0.47}_{-0.30}$ & $6.57^{+1.03}_{-0.73}$ & $6.75^{+0.66}_{-0.42}$ & $9.20^{+1.04}_{-0.66}$\\
 & 3T & & & $5.46\pm0.61$ & $7.43^{+1.95}_{-1.62}$ & $7.65\pm0.85$ & $10.4^{+2.7}_{-2.3}$\\
 & 1P+1T & & & $5.64^{+0.52}_{-0.43}$ & $7.76^{+1.43}_{-1.59}$ & $7.89^{+0.73}_{-0.60}$ & $10.9^{+2.0}_{-2.2}$\\
 & O*2T & & & $4.86^{+0.53}_{-0.35}$ & $110^{+200}_{68}$ & $6.80^{+0.74}_{-0.49}$ & $570^{+280}_{-97}$ \\
\hline \hline
\multicolumn{8}{p{6.0in}}{\T Notes: Fluxes and luminosities are calculated in the $0.3-2$ keV band.  Distance=170 Mpc.  Errors on observed properties are 1$\sigma$, on intrinsic are 90\%.  Model abbreviations: P-photoionization component, O-obscuration in excess of Galactic $N_{\mathrm{H}}$ associated with Mrk 78, T-thermal component.
} \\
\end{tabular}
\label{tab:fluxlum}
\end{minipage}
\end{table*}

\section{Discussion}
\label{sec:discussion}
\subsection{Obscuration within Mrk 78 host galaxy}

Based on the models described in \S\ref{sec:spectral}, the total observed $0.3-2$ keV luminosity arising from thermal or photoionized emission in the nuclear and E-W biconical regions of Mrk 78 ranges from $1.1-1.4\times10^{41}$ erg s$^{-1}$, while the intrinsic $0.3-2$ keV luminosity ranges from $1.5-63\times10^{41}$ erg s$^{-1}$.  Including emission in the N-S cross-conical regions increases these estimates by 10\%.  

\par
The large range in measurements of the intrinsic luminosity arises from uncertainty about whether significant obscuring material within the host galaxy exists.  When we allow obscuration associated with the host galaxy to be a free parameter in the spectral models, we find best-fit values of 7.6, 4.5, and 3.5 $\times10^{21}$ cm$^{-2}$ for the nuclear, inner, and outer regions, respectively.  \citetalias{fischer11} measured a reddening of $E(B-V)=0.78$ for the nuclear region of Mrk 78 based on the $H\alpha/H\beta$ ratio from \textit{HST} STIS spectra.  Using the conversion factor from \citet{savage79}, this reddening corresponds to $4.5\times10^{21}$ cm$^{-2}$.  Thus, the $N_{\mathrm{H}}$ values measured are reasonable, but since the area covered by the STIS slits and the area of X-ray emission are not the same, we cannot be certain that adopting these $N_{\mathrm{H}}$ values in the X-ray spectral fits is the most appropriate choice.  A reddening map of Mrk 78 would allow us to determine appropriate values of $N_{\mathrm{H}}$ in different regions, breaking the degeneracy between obscured and unobscured spectral models, and obtaining more accurate measurements of the intrinsic luminosity. \par
The column densities observed in the central few kiloparsecs of nearby star-forming galaxies are observed to be $N_H\sim10^{21}-10^{22}$ cm$^{-2}$ (e.g. \citealt{mineo12a}; \citealt{mineo12b}; \citealt{kahre18}).  The obscuring columns required by the models for Mrk 78 fall in that range. 

\subsection{Multi-wavelength overview of Mrk 78 emission}
Combined with previous studies of the optical gas ionization, morphology, and kinematics (\citetalias{whittle04}; \citetalias{whittle05}; \citetalias{fischer11}; \citetalias{revalski21}), our X-ray spectral analysis paints a picture in which nuclear radiation is responsible for ionizing the bulk of the [OIII] gas and a significant fraction of the X-ray emitting gas.  Both the X-ray and optical gas may be radiatively driven outwards, although it is also possible that the radio flow plays a role in accelerating these phases and sweeping them into an expanding bubble on the Western side as proposed by \citetalias{whittle04}.  In the Western arc, the outflow likely runs into denser gas; compression of the hot gas boosts its emissivity and may partly responsible for the observed bright arc of X-ray emission. However, as discussed in the next section, there is also evidence that some shocked emission exists at bright X-ray knots on both the Eastern and Western sides.  \par
These bright X-ray knots (XE-1, XW-1, XW-2) coincide with regions where both the optical and X-ray gas appear to play a role in channeling the flow of the radio source.  At the Eastern knot, XE-1, which is bright both in [OIII] and soft X-rays, the radio flow is deflected and changes direction; a knot of enhanced 900 eV emission (XE900-1) likely associated with Ne IX is located just north of the radio knot.  Furthermore, the [OIII] line profile at this knot shows a clear split, which may indicate lateral expansion of the [OIII] gas away from the jet axis \citepalias{whittle04}, and the 900 eV emission, which may arise from shocks, is radially exterior to the [OIII] peak.  These observations point to very dense gas at the location of the Eastern knot. \par
In contrast, on the Western side, the outflow is able to expand outwards, likely running into a more extended gas structure farther out from the nucleus at the location of the Western arc.  Within the Western arc, bright X-ray (XW-1, XW-2) and [OIII] knots are visible where the radio flow appears pinched; these bright X-ray knots also exhibit enhanced 900 eV emission (XW900-1, XW900-2).  Both on the Eastern and Western sides, the bright X-ray knots exhibiting enhanced 900 eV emission indicative of shocks coincide with locations of extreme [OIII] velocity changes of $\sim1000$ km s$^{-1}$.

\subsection{Locations of possible shocked emission}
\label{sec:shocklocation}
A key question which motivated these \textit{Chandra} observations was whether the deceleration of the [OIII] outflow in Mrk 78 is associated with termination shocks.  Our spectral analysis indicates that the soft X-ray emission of Mrk 78 arises from a complex multi-phase medium as evidenced by the fact that a minimum of three model components are required to fit the $0.3-2$ keV extended emission between $1-4.5^{\prime\prime}$ from the nucleus (\S\ref{sec:spectralext}).  Since the only three-component combination that is strongly disfavored for the extended region is a mixture of three photoionization models, it is likely that at least some shocked emission is present. However, it is important to determine the locations where this shocked emission originates to understand its potential connection to the [OIII] outflow.\par
While models consisting of only thermal components are disfavored in the outer region, both the inner and outer regions can be well fit by photoionization-only or mixed photoionization plus thermal models.  Thus, the spectra of these individual regions do not help to pinpoint where the shocked emission may arise.  The only clear difference between the inner and outer regions is that the spectrum of the outer region is overall softer, which can be explained by either lower obscuration and/or an average lower low $kT$ or $U$. \par
One indicator of shocked emission is enhanced NeIX emission with a rest-frame energy of 905 eV (\citealt{paggi12};\citealt{maksym19}; \citealt{paggi22}).  The strongest enhancements of 900 eV emission are found in the Eastern knot (XE900-1) and the Western arc (XW900-1 and XW900-2).  Both of these locations are coincident with a large decrease in [OIII] velocities ($\sim1000$ km s$^{-1}$) as measured by the STIS slits.  The Eastern knot also coincides with the location of the [OIII] outflow turnover radius as modeled by \citetalias{fischer11} and \citetalias{revalski21}.  However, on the Western side, at the location of the model turnover radius and the point at which the radio flow begins to widen outwards, there is not a significant excess of 900 eV emission. In fact, this location exhibits only faint X-ray emission, indicating the gas density in this region may be low.\par
Another indicator of shocked emission are lower $L$[OIII]/$L_X$ ratios than expected for photoionized emission.  Since our spectral modeling shows that the ionization parameter either decreases or remains constant with radius, if all the X-ray emission results from photoinization, the $L$[OIII]/$L_X$ ratio should either remain constant or increase with radius \citep{bianchi06}.  The Eastern knot coincides with the maximum $L$[OIII]/$L_X$ values observed, and therefore, this ratio could be consistent with photoionized emission. However, on the Western side, the $L$[OIII]/$L_X$ ratio decreases with distance from the nucleus, and is especially low in the Western arc.  This trend is inconsistent with the assumption that all X-ray emission in the Western arc is due to photoionization.\par
The Eastern knot (XE-1) and the brightest spots of X-ray emission in the Western arc (XW-1 and XW-2), all of which coincide or are adjacent to bright 900 eV emission, also share similar radio morphology.  The radio emission appears pinched in the Western arc and pinched and deflected at the Eastern knot, suggesting some significant interaction between the radio-emitting flow and the gas at these locations.  Thus, shocks may be present at both of these locations.

\subsection{Shock Energetics and Timescales}
\label{sec:shocks}
As discussed in \S\ref{sec:intro}, the [OIII] outflow may be decelerated by termination shocks, which we expect to inject thermal energy in the gas and produce thermal X-ray emission.  Here we compare the energetics of the [OIII] outflow with those of the possible shocked emission regions.  \par
We first estimate the kinetic power loss associated with the deceleration of the outflow as modeled by \citetalias{fischer11}. The kinetic power is given by $L_{\mathrm{KE}}=0.5\dot M v_{\mathrm{max}}^2$, where $\dot M$ is the mass outflow rate at $r_t$.  The NLR mass can be estimated as $M_{\mathrm{NLR}}/M_{\odot}=7\times10^5 L_{41}(H\beta)/n_3$, where $L_{41}(H\beta)$ is the H$\beta$ luminosity in units of $10^{41}$ erg s$^{-1}$ and $n_3$ is the electron density in units of $10^3$ cm$^{-3}$ \citep{peterson97}.  The H$\beta$ flux of the NLR in a 10$^{\prime\prime}$-radius aperture is $5.0\times10^{-14}$ erg cm$^{-2}$ s$^{-1}$ \citep{mulchaey94}.  Based on the observed Balmer decrement of $f(H\alpha)/f(H\beta)=6.46$ \citep{mulchaey94}, the extinction corrected H$\beta$ luminosity using the \citet{calzetti00} dust law is $1.5\times10^{42}$ erg s$^{-1}$; the H$\beta$ luminosity associated with the central $3^{\prime\prime}$-radius region is only a fraction of this total due to the large aperture used for the H$\beta$ measurement.  Assuming $n_3$=1 based on SII-derived measurements by \citetalias{whittle05}, the NLR mass is $M_{\mathrm{NLR}}=1.1\times10^7 M_{\odot}$.  The time it takes for the outflow to reach its maximum velocity of $v_{\mathrm{max}}=1200$ km s$^{-1}$ at the turnover radius of 700 pc can be found by integrating $\int\frac{dr}{v}$ from the \citetalias{fischer11} outflow model, which gives an estimate of 4 Myr. Thus, the NLR mass outflow rate is $\lesssim2.8 M_{\odot}$ year$^{-1}$, and $L_{\mathrm{KE}}\lesssim1.3\times10^{42}$ erg s$^{-1}$. \par
We can compare the estimated kinetic power loss of the [OIII] outflow with the X-ray luminosity and the thermal power associated with the possible shocks in the Western arc and the Eastern knot. 
For these estimates, we extract spectra from the Western arc and the Eastern knot.  We fit these spectra independently, using the models that include thermal components from Table \ref{tab:spectra} for the outer and inner spectral regions, respectively, fixing all parameters to the best-fit values except the component normalizations.  The $0.3-2$ keV luminosity, corrected for absorption, associated with the thermal spectral components of the Western arc is $L_{\mathrm{apec}}\approx0.04-1.4\times10^{42}$ erg s$^{-1}$.
For the Eastern knot, the $0.3-2$ keV luminosity is $L_{\mathrm{apec}}\approx1-5\times10^{40}$ erg s$^{-1}$.  \par
As an example of our estimates of the thermal power injected into the interstellar medium by shocks, we consider the model with one photoionization and one thermal component with $kT=0.9$ keV from Table \ref{tab:spectra} applied to the Western arc bright spots.  Fitting this model to the Western spectrum, we find an \texttt{apec} normalization of $5\pm2\times10^{-6}$.  This normalization is proportional to the emission measure: $\frac{10^{-14}}{4\pi [D_A(1+z)]^2}\int n_{\mathrm{e}} n_{\mathrm{H}} dV$, giving $n_{\mathrm{e}}=0.3$ cm$^{-3}$, assuming $n_{\mathrm{e}}\approx n_{\mathrm{H}}$.  The thermal pressure is then $p_{\mathrm{th}}\sim2n_{\mathrm{e}}kT$, and the thermal energy is $E_{\mathrm{th}}=p_{\mathrm{th}}V$, where V is the emitting volume.  Assuming that the size of the Western loop along the line-of-sight is comparable to its dimension in the plane of the sky, $1\farcs2$, the total volume of the bright region of the Western arc is 0.3 arcsec$^3=1.6\times10^{64}$ cm$^{3}$. The thermal pressure is then $p_{\mathrm{th}}=1.7\times10^{-9}$ dyne cm$^{-2}$, and the thermal energy is $E_{\mathrm{th}}=1.4\times10^{55}$ erg.  The cooling time of the shocked material is $\tau_c=E_{\mathrm{th}}/L_{\mathrm{apec}}=30$ Myr. \par
The thermal power of the shocks can be derived by estimating the crossing time across the Western arc, which is about $1\farcs2$ wide in the E-W direction.  The thermal velocity of the $0.9$ keV gas is $v_{\mathrm{th}}\sim\sqrt{kT/m_H}\sim290$ km s$^{-1}$, which is comparable to the 200 km s$^{-1}$ velocity dispersion of the [OIII] gas in this region.  Thus, the crossing time is $\approx3$ Myr, and the thermal power of the shocked region is $L_{\mathrm{th}}=E_{\mathrm{th}}/t_{\mathrm{cross}}\approx1.6\times10^{41}$ erg s$^{-1}$.  Deriving $L_{\mathrm{th}}$ estimates using the other thermal models in Table \ref{tab:spectra} normalized to the knot spectrum produces values up to a factor of 3 higher than this value ($L_{\mathrm{th}}=5\times10^{41}$ erg s$^{-1}$).  Performing the same calculation for the Eastern knot, we estimate the thermal power injected by shocks in this region is $L_{\mathrm{th}}=3-6\times10^{40}$ erg s$^{-1}$.
\par
The total amount of power released either as X-ray emission or thermal motion by the shocks that are likely present in the Western arc and Eastern knot is $0.2-2\times10^{42}$ erg s$^{-1}$.  The lower end of this range is based on spectral models with one thermal and one photoionized component, while the higher end assumes that all the soft X-ray emission in these regions has a thermal origin and is significantly obscured.  Therefore, the kinetic power lost by the [OIII] outflow may be accounted for by the total amount of power released by the shocks.
\par

\subsubsection{On comparing locations of shocked X-ray emission with the [OIII] biconical outflow model}
As noted in \S\ref{sec:shocklocation}, while there is evidence suggestive of shocks in the Eastern knot, at the location of the [OIII] outflow turnover radius of $r_t\approx700-900$ pc from the \citetalias{fischer11} and \citetalias{revalski21} bicone models, there is a dearth of X-ray emission 700 pc West of the nucleus.  However, significant changes can be observed in the radio outflow about 700 pc West of the nucleus, as the radio flow widens, and we do find evidence for shocks farther out in the Western arc, at roughly 1700 pc from the nucleus. \par

We consider four plausible explanations for the lack of X-ray emission at the modeled [OIII] turnover radius on the Western side of the outflow:
\par
(1) Given the asymmetric morphology of the outflow on the East and West sides,  the [OIII] turnover radius derived from the symmetric biconical models of \citetalias{fischer11} and \citetalias{revalski21} may not be appropriate for the Western side.  These models in fact do not fit the [OIII] outflow kinematics on the Western side as well as on the Eastern side, although it remains unclear whether this discrepancy may be due to obscuration by dust \citepalias{fischer11}. Thus, one possibility is that there is no significant bulk deceleration of the outflow $700-900$ pc West of the nucleus, and therefore no X-ray emission arising from shocks near this location.  The lack of photoionized X-ray emission in this region may be explained by low gas density and thus correspondingly low X-ray emissivity. \par
(2) Alternatively, bulk deceleration of the [OIII] outflow may occur as on the Eastern side, but both the [OIII] emission and soft X-ray emission may be obscured.  However, more thorough multi-wavelength imaging would be required to evaluate this possibility and determine whether the Western side of the outflow lies behind or in front of the host galaxy disk (\citetalias{fischer11}).
\par
(3) Another possibility is that shocks may occur near the location of the modeled [OIII] turnover radius on the Western side, but they may heat gas to such high temperatures that the gas cannot cool efficiently and thus X-ray emission is only seen farther out in the Western arc once the shock-heated gas has cooled.  As in Herbig-Haro object jets, the shocks may heat gas to temperatures $>$2 keV which results in a longer cooling time due to the lack of line emission at these high temperatures. In this case the cooling distance, $d_{cool}$ can be large. \citet{hartigan87}, \citet{raga02}, and \citet{heathcote98} give $d_{cool}=\frac{100 \mathrm{~cm}^{-3}}{n_0}\frac{v_{shock}}{100\mathrm{~km~s}^{-1}}$. For typical hot-phase ISM values of $n_0 = 1\mathrm{~cm}^{-3}$ and $v_{shock} = 1000\mathrm{~km~s}^{-1}$, $d_{cool} = 200\mathrm{~pc}$ ($\sim0\farcs3$.)
\par
The cooling length formula used for H-H objects is an approximation based on a power law approximation to the radiative cooling coefficient that is appropriate for speeds of a few hundred km~s$^{-1}$. A better approximation for higher speeds is $d_{cool}~ = ~1/4~v_s~t_{cool}$ (John Raymond, 2020, private communication). For  a 1200 km s$^{-1}$ shock, $T \sim 2\times 10^{7}$~K and $\Lambda = 3\times 10^{-23}\mathrm{~erg~cm}^3\mathrm{~s}^{-1}$. Then $t_{cool} = 5/2 (n_e+n_p) kT / (n_e n_p \Lambda)$ and $n_e~=~n_p \sim n_0$ gives $\sim$3~Myr and $d_{cool}~\sim$ 1~kpc ($\sim1\farcs4$).  
\par
At the point where the gas cools sufficiently for strong line emission to occur, the volumetric cooling rate ($\Lambda$) will be up to an order of magnitude larger, and so cooling proportionately more rapid \citep{gnat07}, leading to a sudden brightening and an apparently disconnected arc of X-ray emission. Given these values for $d_{cool}$ it is plausible that the Western arc of emission in Mrk 78 could signal the delayed release of the energy dissipated in a shock at the [OIII] deceleration point as modeled by \citetalias{fischer11}.  However, it is not clear why this ``slow cooling" scenario may occur on the Western side but not on the Eastern side of Mrk 78, where there is a bright knot of X-ray emission which may partly arise from shocks at the location of the [OIII] turnover radius.  
One possibility is that the density of the medium that the outflow impacts is higher on the Eastern side, resulting in a shorter cooling time. \par
(4) Finally, we note that although \citetalias{revalski21} model the [OIII] kinematics as a biconical outflow with a deprojected turnover radius of 900 pc, their photoionization modeling of multiple optical emission lines suggests that the kinetic energy profile of the outflow does not decrease at this radius, but rather increases out to about 1.3 kpc, and remains fairly constant out to 2.3 kpc.  The kinetic energy and mass outflow rate profiles they derive are consistent with in situ acceleration of gas rather than a steady nuclear flow.  If true, then the observed X-ray shocks only dissipate a small fraction of the total kinetic energy of the outflow.
However, the analysis in \citetalias{revalski21} is one-dimensional, treating the Eastern and Western sides as symmetric.  Given the substantial differences between the multi-wavelength emission on the Eastern and Western sides of the outflow, a more spatially resolved analysis of the multi-phase outflow in Mrk 78 may be important for a better understanding of the outflow energetics.

\subsection{AGN Feedback Efficiency}
Assuming the X-ray knots in the Western arc and the Eastern knot do arise from shocks, we can estimate the efficiency with which the AGN injects energy into the ISM via shocks.  In this case, the ratio of the thermal power associated with the knots ($L_{\mathrm{th}}$) and the AGN bolometric luminosity provides a measure of the AGN feedback efficiency.  The total thermal power associated with shocks was found to be $L_{\mathrm{th}}=2-6\times10^{41}$ erg s$^{-1}$.  One way to estimate the bolometric luminosity is to apply a bolometric correction to the intrinsic X-ray luminosity.  The intrinsic $2-10$ keV luminosity of Mrk 78 is measured to be approximately $1\times10^{43}$ erg s$^{-1}$ \citep{zhao20}.  We estimate its bolometric luminosity by adopting a bolometric correction factor of 10 (\citealt{marconi04}; \citealt{lusso12}; \citealt{duras20}), which gives $L_{\mathrm{bol}}\approx1\times10^{44}$ erg s$^{-1}=2.5\times10^{10} L_{\odot}$.  An alternative estimate of the bolometric luminosity can be derived from the far-infrared radiation, which is dominated by the AGN based on its WISE colors (see \S\ref{sec:intro}).  Following \citet{sanders96}, \citetalias{whittle05} estimate the infrared luminosity based on \textit{IRAS} fluxes to be $L_{\mathrm{IR}}\sim4\times10^{44}$ erg s$^{-1}$.  This $L_{\mathrm{IR}}$ estimate is expected to be comparable to the bolometric luminosity for IR AGN like Mrk 78.  
\par
Thus, considering the possible range of $L_{\mathrm{th}}$ and $L_{\mathrm{bol}}$, we calculate that $0.05-0.6$\% of the AGN bolometric luminosity is converted into thermal energy that can heat the ISM.  These estimates are lower than the canonical 5\% required by theoretical models of efficient AGN feedback that can shut down star formation (e.g., \citealt{dimatteo05}; \citealt{hopkins06}), but may reach the 0.5\% estimate of \citet{hopkins10}.  Thus, the thermal energy associated with shocks may be sufficient to shut down star formation, but the large uncertainties in the $L_{\mathrm{th}}$ and $L_{\mathrm{bol}}$ estimates do not allow a firm conclusion to be reached. 
\par

\section{Conclusions}

We have imaged the inner kpc of the type 2 AGN Mrk 78 at sub-arcsecond resolution in X-rays with {\em Chandra} ACIS and find a complex morphology with spectral variations. The overall E-W extent follows approximately that of the optical bi-cone ([OIII]) and the radio (3.6 cm). The Eastern side shows a compact ($\sim$ 700 pc diameter) knot of X-rays coincident with the radio knot.  This knot exhibits a high $L$[OIII]/$L_X$ ratio, consistent with photoionization, but it also shows enhanced 900 eV emission likely associated with Ne IX which is indicative of shocks. The Western side is quite different, being dominated by an extended loop of X-ray emission $\sim$1.7 kpc from the nucleus and $\sim$1.4 kpc in diameter.  This Western arc exhibits regions of $L$[OIII]/$L_X$ ratios that are lower than expected based on our photoionization models which indicate that the photoionization parameter decreases with radius.  These low $L$[OIII]/$L_X$ ratios and the enhanced Ne IX emission in the Western arc are likely indicative of shocks.\par
Spectrally, within 1$^{\prime\prime}$ of the nucleus we find the typical hard spectrum plus neutral Fe-K line of obscured AGN, with a possible detection of another 5.2 keV emission line that could be ascribed to vanadium, a spallation product. In the extended emission regions from $1^{\prime\prime}-4\farcs5$ we find complex spectra requiring at least two components, either photoionized or thermal, and possible intrinsic obscuration (N$_H\sim 10^{21}\mathrm{~cm}^{-2}$).  Spectral fitting of the extended emission overall prefers models which include thermal models representative of shocked emission over models that only include photoionization.  
\par
The intrinsic X-ray luminosity associated with the thermal spectral components of the knots which exhibit evidence for shocked emission is $L_X\approx0.05-1.4\times10^{42}$ erg s$^{-1}$.  We estimate that the thermal energy that may be injected into the interstellar medium by these shocks is $L_{\mathrm{th}}\approx2-6\times10^{41}$ erg s$^{-1}$.  Thus, the total power released by the shocks in these regions is estimated to be $0.2-2\times10^{42}$ erg s$^{-1}$. The power released by the shocks could account for the kinetic power lost by the deceleration of the [OIII] biconical outflow, as modeled by \citetalias{fischer11}.  However, recent modeling of the optical outflow kinematics and photoionization by \citetalias{revalski21} suggests that the kinetic energy profile may actually continue increasing out to 1.3 kpc, and that the overall kinetic energy associated with the outflow may be greater than our estimates due to in situ acceleration of gas out to large radii. \par
The thermal power injected into the interstellar medium constitutes 0.05-0.6\% of the AGN bolometric luminosity.  The upper range of these thermal power estimates may be sufficient to effectively shut down star formation \citep{hopkins10}, but the large uncertainty prevents us from reaching a firm conclusion.\par
On the Eastern side, the location of the soft X-ray knot in which shocks may occur coincides with the location at which the radio flow is deflected and where a large spread in [OIII] velocities is observed; the \citetalias{fischer11} and \citetalias{revalski21} outflow models identify this location $700-900$ pc East of the nucleus as the turnover radius where the [OIII] outflow begins to decelerate.  On the Western side, the bright X-ray knots in the Western arc where shocks are likely to occur coincide with the radio flow being pinched and a large drop in [OIII] velocities.  However, these X-ray knots lie about 1700 pc West of the nucleus, farther out than the turnover radius of the [OIII] outflow model.  This discrepancy between the location of the modeled turnover radius and the shocks on the Western side may indicate that the assumption of a symmetric outflow by \citetalias{fischer11} and \citetalias{revalski21} is not appropriate for Mrk 78, that soft X-ray shocked emission closer to the nucleus is present but strongly absorbed, or that shocked gas closer to the nucleus may be too hot to cool rapidly, leading to an offset of $\sim$1 kpc of the Western X-ray emission from the outflow turnover radius.  If such offsets between the modeled turnover radius and shocked emission are common, \textit{Chandra} may be able to resolve the shocked emission of a greater number of nearby AGN. 
\par
A better understanding of the origin of the X-ray emission in Mrk 78 and its relationship to the biconical outflow requires stronger spectral constraints.  Robustly testing whether the spectrum of the X-ray knots is more consistent with thermal shock emission rather than photoionization would require prohibitive exposure times with current facilities (100 ks of \textit{Chandra} only yields $\approx200$ counts in the $0.3-2$ keV band within the Eastern knot and the bright knots in the Western arc, combined). If instead slow cooling is the correct explanation, then there should be a weak bremsstrahlung emission, below the detection threshold for {\em Chandra}, in the region between the [OIII] deceleration point and the X-ray bright Western arc. Both hypotheses would require an X-ray telescope with subarcsecond spatial resolution and $>10$ higher sensitivity. 

\bigskip

We thank the anonymous reviewer for their feedback to improve the clarity and presentation of the paper.  Support for this work was provided by the National Aeronautics and Space Administration through Chandra Award Number GO6-17084X issued by the Chandra X-ray Center, which is operated by the Smithsonian Astrophysical Observatory for and on behalf of the National Aeronautics Space Administration under contract NAS8-03060.  We thank John Raymond for discussions on cooling times. The scientific results reported in this article are based on observations made by (1) the \textit{Chandra X-ray Observatory}, (2) the NASA/ESA Hubble Space Telescope, obtained from the data archive at the Space Telescope Science Institute, and (3) the Karl G. Jansky Very Large Array, operated by the National Radio Astronomy Observatory (NRAO). STScI is operated by the Association of Universities for Research in Astronomy, Inc. under NASA contract NAS 5-26555. The National Radio Astronomy Observatory is a facility of the National Science Foundation operated under cooperative agreement by Associated Universities, Inc. This work made use of software provided by the \textit{Chandra} X-ray Center (CXC) in the CIAO application package.
\par
\textit{Software:} CIAO \citep{fruscione06}, XSPEC \citep{arnaud96}, CLOUDY \citep{ferland98}
\par
\textit{Facilities:} CXO, HST, VLA, WISE

\bibliography{refs}{}

\begin{thebibliography}{}
\expandafter\ifx\csname natexlab\endcsname\relax\def\natexlab#1{#1}\fi
\providecommand{\url}[1]{\href{#1}{#1}}
\providecommand{\dodoi}[1]{doi:~\href{http://doi.org/#1}{\nolinkurl{#1}}}
\providecommand{\doeprint}[1]{\href{http://ascl.net/#1}{\nolinkurl{http://ascl.net/#1}}}
\providecommand{\doarXiv}[1]{\href{https://arxiv.org/abs/#1}{\nolinkurl{https://arxiv.org/abs/#1}}}

\bibitem[{{Alatalo} {et~al.}(2015){Alatalo}, {Lacy}, {Lanz}, {Bitsakis},
  {Appleton}, {Nyland}, {Cales}, {Chang}, {Davis}, {de Zeeuw}, {Lonsdale},
  {Mart{\'\i}n}, {Meier}, \& {Ogle}}]{alatalo15}
{Alatalo}, K., {Lacy}, M., {Lanz}, L., {et~al.} 2015, \apj, 798, 31,
  \dodoi{10.1088/0004-637X/798/1/31}

\bibitem[{{Anders} \& {Grevesse}(1989)}]{anders89}
{Anders}, E., \& {Grevesse}, N. 1989, \gca, 53, 197,
  \dodoi{10.1016/0016-7037(89)90286-X}

\bibitem[{{Arnaud}(1996)}]{arnaud96}
{Arnaud}, K.~A. 1996, in Astronomical Society of the Pacific Conference Series,
  Vol. 101, Astronomical Data Analysis Software and Systems V, ed. G.~H.
  {Jacoby} \& J.~{Barnes}, 17

\bibitem[{{Bianchi} {et~al.}(2006){Bianchi}, {Guainazzi}, \&
  {Chiaberge}}]{bianchi06}
{Bianchi}, S., {Guainazzi}, M., \& {Chiaberge}, M. 2006, \aap, 448, 499,
  \dodoi{10.1051/0004-6361:20054091}

\bibitem[{{Calzetti} {et~al.}(2000){Calzetti}, {Armus}, {Bohlin}, {Kinney},
  {Koornneef}, \& {Storchi-Bergmann}}]{calzetti00}
{Calzetti}, D., {Armus}, L., {Bohlin}, R.~C., {et~al.} 2000, \apj, 533, 682,
  \dodoi{10.1086/308692}

\bibitem[{{Cid Fernandes} {et~al.}(2001){Cid Fernandes}, {Heckman}, {Schmitt},
  {Gonz{\'a}lez Delgado}, \& {Storchi-Bergmann}}]{fernandes01}
{Cid Fernandes}, R., {Heckman}, T., {Schmitt}, H., {Gonz{\'a}lez Delgado},
  R.~M., \& {Storchi-Bergmann}, T. 2001, \apj, 558, 81, \dodoi{10.1086/322449}

\bibitem[{{Clements}(1981)}]{clements81}
{Clements}, E.~D. 1981, \mnras, 197, 829, \dodoi{10.1093/mnras/197.4.829}

\bibitem[{{Crenshaw} {et~al.}(2015){Crenshaw}, {Fischer}, {Kraemer}, \&
  {Schmitt}}]{crenshaw15}
{Crenshaw}, D.~M., {Fischer}, T.~C., {Kraemer}, S.~B., \& {Schmitt}, H.~R.
  2015, \apj, 799, 83, \dodoi{10.1088/0004-637X/799/1/83}

\bibitem[{{Cutri} \& {et al.}(2013)}]{cutri13}
{Cutri}, R.~M., \& {et al.} 2013, VizieR Online Data Catalog, II/328

\bibitem[{{Das} {et~al.}(2006){Das}, {Crenshaw}, {Kraemer}, \& {Deo}}]{das06}
{Das}, V., {Crenshaw}, D.~M., {Kraemer}, S.~B., \& {Deo}, R.~P. 2006, \aj, 132,
  620, \dodoi{10.1086/504899}

\bibitem[{{de Bruyn} \& {Sargent}(1978)}]{debruyn78}
{de Bruyn}, A.~G., \& {Sargent}, W.~L.~W. 1978, \aj, 83, 1257,
  \dodoi{10.1086/112320}

\bibitem[{{Di Matteo} {et~al.}(2005){Di Matteo}, {Springel}, \&
  {Hernquist}}]{dimatteo05}
{Di Matteo}, T., {Springel}, V., \& {Hernquist}, L. 2005, \nat, 433, 604,
  \dodoi{10.1038/nature03335}

\bibitem[{{Duras} {et~al.}(2020){Duras}, {Bongiorno}, {Ricci}, {Piconcelli},
  {Shankar}, {Lusso}, {Bianchi}, {Fiore}, {Maiolino}, {Marconi}, {Onori},
  {Sani}, {Schneider}, {Vignali}, \& {La Franca}}]{duras20}
{Duras}, F., {Bongiorno}, A., {Ricci}, F., {et~al.} 2020, \aap, 636, A73,
  \dodoi{10.1051/0004-6361/201936817}

\bibitem[{{Dutson} {et~al.}(2014){Dutson}, {Edge}, {Hinton}, {Hogan},
  {Gurwell}, \& {Alston}}]{dutson14}
{Dutson}, K.~L., {Edge}, A.~C., {Hinton}, J.~A., {et~al.} 2014, \mnras, 442,
  2048, \dodoi{10.1093/mnras/stu975}

\bibitem[{{Elvis} {et~al.}(1994){Elvis}, {Wilkes}, {McDowell}, {Green},
  {Bechtold}, {Willner}, {Oey}, {Polomski}, \& {Cutri}}]{elvis94}
{Elvis}, M., {Wilkes}, B.~J., {McDowell}, J.~C., {et~al.} 1994, \apjs, 95, 1,
  \dodoi{10.1086/192093}

\bibitem[{{Ferland} {et~al.}(1998){Ferland}, {Korista}, {Verner}, {Ferguson},
  {Kingdon}, \& {Verner}}]{ferland98}
{Ferland}, G.~J., {Korista}, K.~T., {Verner}, D.~A., {et~al.} 1998, \pasp, 110,
  761, \dodoi{10.1086/316190}

\bibitem[{{Ferland} {et~al.}(2017){Ferland}, {Chatzikos}, {Guzm{\'a}n},
  {Lykins}, {van Hoof}, {Williams}, {Abel}, {Badnell}, {Keenan}, {Porter}, \&
  {Stancil}}]{ferland17}
{Ferland}, G.~J., {Chatzikos}, M., {Guzm{\'a}n}, F., {et~al.} 2017, \rmxaa, 53,
  385.
\newblock \doarXiv{1705.10877}

\bibitem[{{Feruglio} {et~al.}(2010){Feruglio}, {Maiolino}, {Piconcelli},
  {Menci}, {Aussel}, {Lamastra}, \& {Fiore}}]{feruglio10}
{Feruglio}, C., {Maiolino}, R., {Piconcelli}, E., {et~al.} 2010, \aap, 518,
  L155, \dodoi{10.1051/0004-6361/201015164}

\bibitem[{{Fiore} {et~al.}(2017){Fiore}, {Feruglio}, {Shankar}, {Bischetti},
  {Bongiorno}, {Brusa}, {Carniani}, {Cicone}, {Duras}, {Lamastra}, {Mainieri},
  {Marconi}, {Menci}, {Maiolino}, {Piconcelli}, {Vietri}, \&
  {Zappacosta}}]{fiore17}
{Fiore}, F., {Feruglio}, C., {Shankar}, F., {et~al.} 2017, \aap, 601, A143,
  \dodoi{10.1051/0004-6361/201629478}

\bibitem[{{Fischer} {et~al.}(2019){Fischer}, {Smith}, {Kraemer}, {Schmitt},
  {Crenshaw}, {Koss}, {Mushotzky}, {Larson}, {U}, \& {Rigby}}]{fischer19}
{Fischer}, T., {Smith}, K.~L., {Kraemer}, S., {et~al.} 2019, \apj, 887, 200,
  \dodoi{10.3847/1538-4357/ab55e3}

\bibitem[{{Fischer} {et~al.}(2013){Fischer}, {Crenshaw}, {Kraemer}, \&
  {Schmitt}}]{fischer13}
{Fischer}, T.~C., {Crenshaw}, D.~M., {Kraemer}, S.~B., \& {Schmitt}, H.~R.
  2013, \apjs, 209, 1, \dodoi{10.1088/0067-0049/209/1/1}

\bibitem[{{Fischer} {et~al.}(2011){Fischer}, {Crenshaw}, {Kraemer}, {Schmitt},
  {Mushotsky}, \& {Dunn}}]{fischer11}
{Fischer}, T.~C., {Crenshaw}, D.~M., {Kraemer}, S.~B., {et~al.} 2011, \apj,
  727, 71, \dodoi{10.1088/0004-637X/727/2/71}

\bibitem[{{Fischer} {et~al.}(2017){Fischer}, {Machuca}, {Diniz}, {Crenshaw},
  {Kraemer}, {Riffel}, {Schmitt}, {Baron}, {Storchi-Bergmann}, {Straughn},
  {Revalski}, \& {Pope}}]{fischer17}
{Fischer}, T.~C., {Machuca}, C., {Diniz}, M.~R., {et~al.} 2017, \apj, 834, 30,
  \dodoi{10.3847/1538-4357/834/1/30}

\bibitem[{{Fischer} {et~al.}(2018){Fischer}, {Kraemer}, {Schmitt}, {Longo
  Micchi}, {Crenshaw}, {Revalski}, {Vestergaard}, {Elvis}, {Gaskell}, {Hamann},
  {Ho}, {Hutchings}, {Mushotzky}, {Netzer}, {Storchi-Bergmann}, {Straughn},
  {Turner}, \& {Ward}}]{fischer18}
{Fischer}, T.~C., {Kraemer}, S.~B., {Schmitt}, H.~R., {et~al.} 2018, \apj, 856,
  102, \dodoi{10.3847/1538-4357/aab03e}

\bibitem[{{Fruscione} {et~al.}(2006){Fruscione}, {McDowell}, {Allen},
  {Brickhouse}, {Burke}, {Davis}, {Durham}, {Elvis}, {Galle}, {Harris},
  {Huenemoerder}, {Houck}, {Ishibashi}, {Karovska}, {Nicastro}, {Noble},
  {Nowak}, {Primini}, {Siemiginowska}, {Smith}, \& {Wise}}]{fruscione06}
{Fruscione}, A., {McDowell}, J.~C., {Allen}, G.~E., {et~al.} 2006, in Society
  of Photo-Optical Instrumentation Engineers (SPIE) Conference Series, Vol.
  6270, \procspie, 62701V, \dodoi{10.1117/12.671760}

\bibitem[{{Gallo} {et~al.}(2019){Gallo}, {Randhawa}, {Waddell}, {Hani},
  {Garc{\'\i}a}, \& {Reynolds}}]{gallo19}
{Gallo}, L.~C., {Randhawa}, J.~S., {Waddell}, S.~G.~H., {et~al.} 2019, \mnras,
  484, 3036, \dodoi{10.1093/mnras/stz260}

\bibitem[{{Gandhi} {et~al.}(2009){Gandhi}, {Horst}, {Smette}, {H{\"o}nig},
  {Comastri}, {Gilli}, {Vignali}, \& {Duschl}}]{gandhi09}
{Gandhi}, P., {Horst}, H., {Smette}, A., {et~al.} 2009, \aap, 502, 457,
  \dodoi{10.1051/0004-6361/200811368}

\bibitem[{{Gnat} \& {Sternberg}(2007)}]{gnat07}
{Gnat}, O., \& {Sternberg}, A. 2007, \apjs, 168, 213, \dodoi{10.1086/509786}

\bibitem[{{Hartigan} {et~al.}(1987){Hartigan}, {Raymond}, \&
  {Hartmann}}]{hartigan87}
{Hartigan}, P., {Raymond}, J., \& {Hartmann}, L. 1987, \apj, 316, 323,
  \dodoi{10.1086/165204}

\bibitem[{{Heathcote} {et~al.}(1998){Heathcote}, {Reipurth}, \&
  {Raga}}]{heathcote98}
{Heathcote}, S., {Reipurth}, B., \& {Raga}, A.~C. 1998, \aj, 116, 1940,
  \dodoi{10.1086/300548}

\bibitem[{{Hlavacek-Larrondo} {et~al.}(2015){Hlavacek-Larrondo}, {McDonald},
  {Benson}, {Forman}, {Allen}, {Bleem}, {Ashby}, {Bocquet}, {Brodwin},
  {Dietrich}, {Jones}, {Liu}, {Reichardt}, {Saliwanchik}, {Saro}, {Schrabback},
  {Song}, {Stalder}, {Vikhlinin}, \& {Zenteno}}]{hlavacek15}
{Hlavacek-Larrondo}, J., {McDonald}, M., {Benson}, B.~A., {et~al.} 2015, \apj,
  805, 35, \dodoi{10.1088/0004-637X/805/1/35}

\bibitem[{{Hopkins} \& {Elvis}(2010)}]{hopkins10}
{Hopkins}, P.~F., \& {Elvis}, M. 2010, \mnras, 401, 7,
  \dodoi{10.1111/j.1365-2966.2009.15643.x}

\bibitem[{{Hopkins} {et~al.}(2006){Hopkins}, {Hernquist}, {Cox}, {Di Matteo},
  {Robertson}, \& {Springel}}]{hopkins06}
{Hopkins}, P.~F., {Hernquist}, L., {Cox}, T.~J., {et~al.} 2006, \apjs, 163, 1,
  \dodoi{10.1086/499298}

\bibitem[{{Ivezi{\'c}} {et~al.}(2002){Ivezi{\'c}}, {Menou}, {Knapp}, {Strauss},
  {Lupton}, {Vand en Berk}, {Richards}, {Tremonti}, {Weinstein}, {Anderson},
  {Bahcall}, {Becker}, {Bernardi}, {Blanton}, {Eisenstein}, {Fan},
  {Finkbeiner}, {Finlator}, {Frieman}, {Gunn}, {Hall}, {Kim}, {Kinkhabwala},
  {Narayanan}, {Rockosi}, {Schlegel}, {Schneider}, {Strateva}, {SubbaRao},
  {Thakar}, {Voges}, {White}, {Yanny}, {Brinkmann}, {Doi}, {Fukugita},
  {Hennessy}, {Munn}, {Nichol}, \& {York}}]{ivezic02}
{Ivezi{\'c}}, {\v{Z}}., {Menou}, K., {Knapp}, G.~R., {et~al.} 2002, \aj, 124,
  2364, \dodoi{10.1086/344069}

\bibitem[{{Jones} {et~al.}(2020){Jones}, {Fabbiano}, {Elvis}, {Paggi},
  {Karovska}, {Maksym}, {Siemiginowska}, \& {Raymond}}]{jones20}
{Jones}, M.~L., {Fabbiano}, G., {Elvis}, M., {et~al.} 2020, \apj, 891, 133,
  \dodoi{10.3847/1538-4357/ab76c8}

\bibitem[{{Kahre} {et~al.}(2018){Kahre}, {Walterbos}, {Kim}, {Thilker},
  {Calzetti}, {Lee}, {Sabbi}, {Ubeda}, {Aloisi}, {Cignoni}, {Cook}, {Dale},
  {Elmegreen}, {Elmegreen}, {Fumagalli}, {Gallagher}, {Gouliermis}, {Grasha},
  {Grebel}, {Hunter}, {Sacchi}, {Smith}, {Tosi}, {Adamo}, {Andrews},
  {Ashworth}, {Bright}, {Brown}, {Chandar}, {Christian}, {de Mink}, {Dobbs},
  {Evans}, {Herrero}, {Johnson}, {Kennicutt}, {Krumholz}, {Messa}, {Nair},
  {Nota}, {Pellerin}, {Ryon}, {Schaerer}, {Shabani}, {Van Dyk}, {Whitmore}, \&
  {Wofford}}]{kahre18}
{Kahre}, L., {Walterbos}, R.~A., {Kim}, H., {et~al.} 2018, \apj, 855, 133,
  \dodoi{10.3847/1538-4357/aab101}

\bibitem[{{Kozlova} {et~al.}(2020){Kozlova}, {Moiseev}, \&
  {Smirnova}}]{kozlova20}
{Kozlova}, D.~V., {Moiseev}, A.~V., \& {Smirnova}, A.~A. 2020, Contributions of
  the Astronomical Observatory Skalnate Pleso, 50, 309,
  \dodoi{10.31577/caosp.2020.50.1.309}

\bibitem[{{Lusso} {et~al.}(2012){Lusso}, {Comastri}, {Simmons}, {Mignoli},
  {Zamorani}, {Vignali}, {Brusa}, {Shankar}, {Lutz}, {Trump}, {Maiolino},
  {Gilli}, {Bolzonella}, {Puccetti}, {Salvato}, {Impey}, {Civano}, {Elvis},
  {Mainieri}, {Silverman}, {Koekemoer}, {Bongiorno}, {Merloni}, {Berta}, {Le
  Floc'h}, {Magnelli}, {Pozzi}, \& {Riguccini}}]{lusso12}
{Lusso}, E., {Comastri}, A., {Simmons}, B.~D., {et~al.} 2012, \mnras, 425, 623,
  \dodoi{10.1111/j.1365-2966.2012.21513.x}

\bibitem[{{Maksym} {et~al.}(2019){Maksym}, {Fabbiano}, {Elvis}, {Karovska},
  {Paggi}, {Raymond}, {Wang}, {Storchi-Bergmann}, \& {Risaliti}}]{maksym19}
{Maksym}, W.~P., {Fabbiano}, G., {Elvis}, M., {et~al.} 2019, \apj, 872, 94,
  \dodoi{10.3847/1538-4357/aaf4f5}

\bibitem[{{Marconi} {et~al.}(2004){Marconi}, {Risaliti}, {Gilli}, {Hunt},
  {Maiolino}, \& {Salvati}}]{marconi04}
{Marconi}, A., {Risaliti}, G., {Gilli}, R., {et~al.} 2004, \mnras, 351, 169,
  \dodoi{10.1111/j.1365-2966.2004.07765.x}

\bibitem[{{Mewe} {et~al.}(1985){Mewe}, {Gronenschild}, \& {van den
  Oord}}]{mewe85}
{Mewe}, R., {Gronenschild}, E.~H.~B.~M., \& {van den Oord}, G.~H.~J. 1985,
  \aaps, 62, 197

\bibitem[{{Michel} \& {Huchra}(1988)}]{michel88}
{Michel}, A., \& {Huchra}, J. 1988, \pasp, 100, 1423, \dodoi{10.1086/132342}

\bibitem[{{Mineo} {et~al.}(2012{\natexlab{a}}){Mineo}, {Gilfanov}, \&
  {Sunyaev}}]{mineo12a}
{Mineo}, S., {Gilfanov}, M., \& {Sunyaev}, R. 2012{\natexlab{a}}, \mnras, 419,
  2095, \dodoi{10.1111/j.1365-2966.2011.19862.x}

\bibitem[{{Mineo} {et~al.}(2012{\natexlab{b}}){Mineo}, {Gilfanov}, \&
  {Sunyaev}}]{mineo12b}
---. 2012{\natexlab{b}}, \mnras, 426, 1870,
  \dodoi{10.1111/j.1365-2966.2012.21831.x}

\bibitem[{{Mulchaey} {et~al.}(1994){Mulchaey}, {Koratkar}, {Ward}, {Wilson},
  {Whittle}, {Antonucci}, {Kinney}, \& {Hurt}}]{mulchaey94}
{Mulchaey}, J.~S., {Koratkar}, A., {Ward}, M.~J., {et~al.} 1994, \apj, 436,
  586, \dodoi{10.1086/174933}

\bibitem[{{Paggi} {et~al.}(2022){Paggi}, {Fabbiano}, {Nardini}, {Karovska},
  {Elvis}, \& {Wang}}]{paggi22}
{Paggi}, A., {Fabbiano}, G., {Nardini}, E., {et~al.} 2022, \apj, 927, 166,
  \dodoi{10.3847/1538-4357/ac5025}

\bibitem[{{Paggi} {et~al.}(2012){Paggi}, {Wang}, {Fabbiano}, {Elvis}, \&
  {Karovska}}]{paggi12}
{Paggi}, A., {Wang}, J., {Fabbiano}, G., {Elvis}, M., \& {Karovska}, M. 2012,
  \apj, 756, 39, \dodoi{10.1088/0004-637X/756/1/39}

\bibitem[{{Peterson}(1997)}]{peterson97}
{Peterson}, B.~M. 1997, {An Introduction to Active Galactic Nuclei}

\bibitem[{{Porquet} \& {Dubau}(2000)}]{porquet00}
{Porquet}, D., \& {Dubau}, J. 2000, \aaps, 143, 495,
  \dodoi{10.1051/aas:2000192}

\bibitem[{{Raga} {et~al.}(2002){Raga}, {Noriega-Crespo}, \&
  {Vel{\'a}zquez}}]{raga02}
{Raga}, A.~C., {Noriega-Crespo}, A., \& {Vel{\'a}zquez}, P.~F. 2002, \apjl,
  576, L149, \dodoi{10.1086/343760}

\bibitem[{{Revalski} {et~al.}(2018{\natexlab{a}}){Revalski}, {Crenshaw},
  {Kraemer}, {Fischer}, {Schmitt}, \& {Machuca}}]{revalski18a}
{Revalski}, M., {Crenshaw}, D.~M., {Kraemer}, S.~B., {et~al.}
  2018{\natexlab{a}}, \apj, 856, 46, \dodoi{10.3847/1538-4357/aab107}

\bibitem[{{Revalski} {et~al.}(2018{\natexlab{b}}){Revalski}, {Dashtamirova},
  {Crenshaw}, {Kraemer}, {Fischer}, {Schmitt}, {Gnilka}, {Schmidt}, {Elvis},
  {Fabbiano}, {Storchi-Bergmann}, {Maksym}, \& {Gandhi}}]{revalski18b}
{Revalski}, M., {Dashtamirova}, D., {Crenshaw}, D.~M., {et~al.}
  2018{\natexlab{b}}, \apj, 867, 88, \dodoi{10.3847/1538-4357/aae3e6}

\bibitem[{{Revalski} {et~al.}(2021){Revalski}, {Meena}, {Martinez}, {Polack},
  {Crenshaw}, {Kraemer}, {Collins}, {Fischer}, {Schmitt}, {Schmidt}, {Maksym},
  \& {Rafelski}}]{revalski21}
{Revalski}, M., {Meena}, B., {Martinez}, F., {et~al.} 2021, \apj, 910, 139,
  \dodoi{10.3847/1538-4357/abdcad}

\bibitem[{{Sanders} \& {Mirabel}(1996)}]{sanders96}
{Sanders}, D.~B., \& {Mirabel}, I.~F. 1996, \araa, 34, 749,
  \dodoi{10.1146/annurev.astro.34.1.749}

\bibitem[{{Savage} \& {Mathis}(1979)}]{savage79}
{Savage}, B.~D., \& {Mathis}, J.~S. 1979, \araa, 17, 73,
  \dodoi{10.1146/annurev.aa.17.090179.000445}

\bibitem[{{Scott}(2005)}]{scott05}
{Scott}, L.~M. 2005, PhD thesis, Washington University, Missouri, USA

\bibitem[{{Skibo}(1997)}]{skibo97}
{Skibo}, J.~G. 1997, \apj, 478, 522, \dodoi{10.1086/303829}

\bibitem[{{Smirnova} {et~al.}(2010){Smirnova}, {Moiseev}, \&
  {Afanasiev}}]{smirnova10}
{Smirnova}, A.~A., {Moiseev}, A.~V., \& {Afanasiev}, V.~L. 2010, \mnras, 408,
  400, \dodoi{10.1111/j.1365-2966.2010.17121.x}

\bibitem[{{Stern} {et~al.}(2012){Stern}, {Assef}, {Benford}, {Blain}, {Cutri},
  {Dey}, {Eisenhardt}, {Griffith}, {Jarrett}, {Lake}, {Masci}, {Petty},
  {Stanford}, {Tsai}, {Wright}, {Yan}, {Harrison}, \& {Madsen}}]{stern12}
{Stern}, D., {Assef}, R.~J., {Benford}, D.~J., {et~al.} 2012, \apj, 753, 30,
  \dodoi{10.1088/0004-637X/753/1/30}

\bibitem[{{Travascio} {et~al.}(2021){Travascio}, {Fabbiano}, {Paggi}, {Elvis},
  {Maksym}, {Morganti}, {Osterloo}, \& {Fiore}}]{travascio21}
{Travascio}, A., {Fabbiano}, G., {Paggi}, A., {et~al.} 2021, \apj

\bibitem[{{Turner} {et~al.}(2010){Turner}, {Miller}, {Reeves}, {Lobban},
  {Braito}, {Kraemer}, \& {Crenshaw}}]{turner10}
{Turner}, T.~J., {Miller}, L., {Reeves}, J.~N., {et~al.} 2010, \apj, 712, 209,
  \dodoi{10.1088/0004-637X/712/1/209}

\bibitem[{{Wang} {et~al.}(2011{\natexlab{a}}){Wang}, {Fabbiano}, {Elvis},
  {Risaliti}, {Mundell}, {Karovska}, \& {Zezas}}]{wang11b}
{Wang}, J., {Fabbiano}, G., {Elvis}, M., {et~al.} 2011{\natexlab{a}}, \apj,
  736, 62, \dodoi{10.1088/0004-637X/736/1/62}

\bibitem[{{Wang} {et~al.}(2012){Wang}, {Fabbiano}, {Karovska}, {Elvis}, \&
  {Risaliti}}]{wang12}
{Wang}, J., {Fabbiano}, G., {Karovska}, M., {Elvis}, M., \& {Risaliti}, G.
  2012, \apj, 756, 180, \dodoi{10.1088/0004-637X/756/2/180}

\bibitem[{{Wang} {et~al.}(2011{\natexlab{b}}){Wang}, {Fabbiano}, {Risaliti},
  {Elvis}, {Karovska}, {Zezas}, {Mundell}, {Dumas}, \& {Schinnerer}}]{wang11a}
{Wang}, J., {Fabbiano}, G., {Risaliti}, G., {et~al.} 2011{\natexlab{b}}, \apj,
  729, 75, \dodoi{10.1088/0004-637X/729/1/75}

\bibitem[{{Wang} {et~al.}(2011{\natexlab{c}}){Wang}, {Fabbiano}, {Elvis},
  {Risaliti}, {Karovska}, {Zezas}, {Mundell}, {Dumas}, \&
  {Schinnerer}}]{wang11c}
{Wang}, J., {Fabbiano}, G., {Elvis}, M., {et~al.} 2011{\natexlab{c}}, \apj,
  742, 23, \dodoi{10.1088/0004-637X/742/1/23}

\bibitem[{{Whittle} {et~al.}(2005){Whittle}, {Rosario}, {Silverman}, {Nelson},
  \& {Wilson}}]{whittle05}
{Whittle}, M., {Rosario}, D.~J., {Silverman}, J.~D., {Nelson}, C.~H., \&
  {Wilson}, A.~S. 2005, \aj, 129, 104, \dodoi{10.1086/426562}

\bibitem[{{Whittle} \& {Wilson}(2004)}]{whittle04}
{Whittle}, M., \& {Wilson}, A.~S. 2004, \aj, 127, 606, \dodoi{10.1086/380940}

\bibitem[{{Whittle} {et~al.}(2002){Whittle}, {Wilson}, {Nelson}, {Rosario}, \&
  {Silverman}}]{whittle02}
{Whittle}, M., {Wilson}, A.~S., {Nelson}, C.~H., {Rosario}, D., \& {Silverman},
  J.~D. 2002, in Revista Mexicana de Astronomia y Astrofisica Conference
  Series, Vol.~13, Revista Mexicana de Astronomia y Astrofisica Conference
  Series, ed. W.~J. {Henney}, W.~{Steffen}, L.~{Binette}, \& A.~{Raga},
  230--235

\bibitem[{{Zhao} {et~al.}(2020){Zhao}, {Marchesi}, {Ajello}, {Balokovi{\'c}},
  \& {Fischer}}]{zhao20}
{Zhao}, X., {Marchesi}, S., {Ajello}, M., {Balokovi{\'c}}, M., \& {Fischer}, T.
  2020, \apj, 894, 71, \dodoi{10.3847/1538-4357/ab879d}

\end{thebibliography}
\bibliographystyle{aasjournal}



\end{document}